\documentclass[lettersize,journal]{IEEEtran}
\usepackage{amsmath,amsfonts}
\usepackage{algorithmic}
\usepackage{algorithm}
\usepackage{array}
\usepackage[caption=false,font=normalsize,labelfont=sf,textfont=sf]{subfig}
\usepackage{textcomp}
\usepackage{stfloats}
\usepackage{url}
\usepackage{verbatim}
\usepackage{color,soul}
\usepackage{graphicx}
\usepackage{hyperref}
\usepackage{cite}
\usepackage[nolist]{acronym}
\usepackage{mathtools}
\usepackage{amssymb}
\usepackage{xfrac}
\acrodefplural{HPA}[HPAs]{high power amplifiers}

\begin{acronym}
	\acro{OFDM}{orthogonal frequency-division multiplexing}
	\acro{DL}{downlink}
	\acro{i.i.d}{independently and identically distributed}
	\acro{UL}{uplink}
	\acro{LoS}{line of sight}
	\acro{DFT}{discrete Fourier transform}
    \acro{HPA}{high power amplifier}
    \acro{IBO}{input power back-off}
    \acro{MIMO}{multiple-input, multiple-output}
    \acro{PAPR}{peak-to-average power ratio}
    \acro{AWGN}{additive white Gaussian noise}
    \acro{KPI}{key performance indicator}
    \acro{KPIs}{key performance indicators}  
    \acro{D2D}{device-to-device}
    \acro{RSU}{roadside unit}
    \acro{ISAC}{integrated sensing and communication}
    \acro{SnC}[S\&C]{sensing and communication}
    \acro{RnC}[R\&C]{radar and communication}
    \acro{DFRC}{dual-functional radar and communication}
    \acro{AoA}{angle of arrival}
    \acro{AoD}{angle of departure}
    \acro{AoAs}{angles of arrival}
    \acro{ToA}{time of arrival}
    \acro{EVM}{error vector magnitude}
    \acro{CPI}{coherent processing interval}
    \acro{eMBB}{enhanced mobile broadband}
    \acro{URLLC}{ultra-reliable low latency communications}
    \acro{mMTC}{massive machine type communications}
	\acro{QCQP}{quadratic constrained quadratic programming}
	\acro{BnB}{branch and bound}
	\acro{SER}{symbol error rate}
	\acro{LFM}{linear frequency modulation}
	\acro{ADMM}{alternating direction method of multipliers}
	\acro{CCDF}{complementary cumulative distribution function}
	\acro{MISO}{multiple-input, single-output}
	\acro{CSI}{channel state information}
	\acro{OTFS}{orthogonal time frequency space}
	\acro{LDPC}{low-density parity-check}
	\acro{BCC}{binary convolutional coding}
	\acro{IEEE}{Institute of Electrical and Electronics Engineers}
	\acro{ULA}{uniform linear antenna}
	\acro{B5G}{beyond 5G}
	\acro{MOOP}{multi-objective optimization problem}
	\acro{ML}{maximum likelihood}
	\acro{FML}{fused maximum likelihood}
	\acro{RF}{radio frequency}
	\acro{AM/AM}{amplitude modulation/amplitude modulation}
	\acro{AM/PM}{amplitude modulation/phase modulation}
	\acro{BS}{base station}
	\acro{MM}{majorization-minimization}
	\acro{LNCA}{$\ell$-norm cyclic algorithm}
	\acro{OCDM}{orthogonal chirp-division multiplexing}
	\acro{TR}{tone reservation}
	\acro{COCS}{consecutive ordered cyclic shifts}
	\acro{LS}{least squares}
	\acro{CVE}{coefficient of variation of envelopes}
	\acro{BSUM}{block successive upper-bound minimization}
	\acro{ICE}{iterative convex enhancement}
	\acro{SOA}{sequential optimization algorithm}
	\acro{BCD}{block coordinate descent}
	\acro{SINR}{signal to interference plus noise ratio}
	\acro{MICF}{modified iterative clipping and filtering}
	\acro{PSL}{peak side-lobe level}
	\acro{SDR}{semi-definite relaxation}
	\acro{KL}{Kullback-Leibler}
	\acro{ADSRP}{alternating direction sequential relaxation programming}
	\acro{MUSIC}{multiple signal classification}
	\acro{EVD}{eigenvalue decomposition}
	\acro{SVD}{singular value decomposition}
	\acro{MSE}{mean squared error}
	\acro{SNR}{signal-to-noise ratio}
	\acro{CRB}{Cram\'er-Rao bound}
	\acro{QPSK}{quadrature phase shift keying}
	\acro{RCS}{radar cross-section}
	\acro{NOMA}{non-orthogonal multiple access}
	\acro{IRS}{intelligent reflecting surface}
	\acro{IoT}{Internet of things}
	\acro{UAV}{unmanned aerial vehicle}
	\acro{V2X}{vehicle-to-everything}
	\acro{SDRs}{sofware-defined radios}
	\acro{FDD}{frequency-division duplex}
	\acro{NR}{new radio}
	\acro{ADCs}{analog-to-digital converters}
	\acro{ADC}{analog-to-digital converter}
	\acro{LNA}{low noise amplifier}
	\acro{GNSS}{global navigation satellite system}
	\acro{NMSE}{normalized mean squared error}
	\acro{HRF}{hybrid radar fusion}
	\acro{mmWave}{millimeter wave}
	\acro{GAMP}{generalized approximate message passing}
	\acro{DPI}{direct path interference}
	\acro{DPIS}{direct path interference suppression}
	\acro{w.r.t}{with respect to}
	\acro{SU}{single-user}
\end{acronym}

\DeclareMathOperator{\Span}{span}
\DeclareMathOperator{\trace}{tr}
\DeclareMathOperator{\SNR}{SNR}

\DeclareMathOperator{\Real}{Re}

\DeclareMathOperator{\dB}{dB}
\DeclareMathOperator{\dBm}{dBm}
\DeclareMathOperator{\opt}{opt}

\DeclareMathOperator{\MSE}{MSE}
\DeclareMathOperator{\rad}{rad}
\DeclareMathOperator{\diag}{diag}
\DeclareMathOperator{\CRB}{CRB}

\DeclareMathOperator*{\argmax}{arg\,max}
\DeclareMathOperator*{\argmin}{arg\,min}

\usepackage{xcolor,cite,etoolbox}
\makeatletter 
\pretocmd\@bibitem{\color{black}\csname keycolor#1\endcsname}{}{\fail}
\newcommand\citecolor[1]{\@namedef{keycolor#1}{\color{blue}}}
\makeatother

\usepackage{fancyhdr}
\fancypagestyle{firststyle}{
	\fancyhf{}
	\fancyhead[L]{A. Chowdary, A. Bazzi, and M. Chafii, ``On Hybrid Radar Fusion for Integrated Sensing and Communication'', accepted in \textit{IEEE Transactions on Wireless Communications}, Jan. 2024.}

}
\begin{document}
\bstctlcite{IEEEexample:BSTcontrol}

\title{On Hybrid Radar Fusion for Integrated Sensing and Communication}

\author{Akhileswar Chowdary Gaddam, Ahmad Bazzi,  Marwa Chafii 
\thanks{
Akhileswar Chowdary Gaddam is with the NYU Tandon School of Engineering, Brooklyn, 11201, NY, USA (email: \href{akhileswar.chowdary@nyu.edu}{akhileswar.chowdary@nyu.edu})

Ahmad Bazzi is with Engineering Division, New York University (NYU) Abu Dhabi, 129188, UAE and NYU WIRELESS,
NYU Tandon School of Engineering, Brooklyn, 11201, NY, USA
(email: \href{ahmad.bazzi@nyu.edu}{ahmad.bazzi@nyu.edu}).

Marwa Chafii is with Engineering Division, New York University (NYU) Abu Dhabi, 129188, UAE and NYU WIRELESS,
NYU Tandon School of Engineering, Brooklyn, 11201, NY, USA (email: \href{marwa.chafii@nyu.edu}{marwa.chafii@nyu.edu}).}
\thanks{Manuscript received xxx}}

\markboth{lorem}%
{Shell \MakeLowercase{\textit{et al.}}: A Sample Article Using IEEEtran.cls for IEEE Journals}

\IEEEpubid{}

\maketitle
\thispagestyle{firststyle}

\begin{abstract}
The following paper introduces a novel integrated sensing and communication (ISAC) scenario termed hybrid radar fusion. In this setting, the dual-functional radar and communications (DFRC) base station (BS) acts as a mono-static radar in the downlink (DL), for sensing purposes, while performing its DL communication tasks. Meanwhile, the communication users act as distributed bi-static radar nodes in the uplink (UL) following a frequency-division duplex protocol. The DFRC BS fuses the information available at different DL and UL resource bands to estimate the angles-of-arrival (AoAs) of the multiple targets existing in the scene. In this work, we derive the maximum likelihood (ML) criterion for the hybrid radar fusion problem at hand. Additionally, we design efficient estimators; the first algorithm is based on an alternating optimization approach to solve the ML criterion, while the second one designs an optimization framework that leads to an alternating subspace approach to estimate AoAs for both the target and users. Finally, we demonstrate the superior performance of both algorithms in different scenarios, and the gains offered by these proposed methods through numerical simulations.
\end{abstract}


\section{Introduction}
\label{sec:introduction}
\IEEEPARstart{D}{\lowercase{espite}} having numerous similarities in terms of signal processing and system design, radar sensing and wireless communication have been evolving individually for many years \cite{9585321}. Interestingly, both \ac{RnC} systems have quite a number of signal processing features that can be shared in common, thus favoring their joint design \cite{8999605,8828030,9705498,10041914,7786130}. Indeed, \ac{ISAC} can open the way for ground-breaking applications, for e.g., in the automotive sector \cite{9830717}, \ac{IoT}, and robotics.


\ac{ISAC} has been investigated for different scenarios and challenges, including
holographic communications \cite{9724245}, 
\ac{DFRC} \cite{9737357},
waveform design \cite{9724187},
\ac{UAV} \cite{10004900},
beamforming design \cite{10018908},
\ac{IoT} \cite{9206051}, just to name a few. Due to the interoperability between the \ac{RnC} sub-systems, \ac{ISAC} has a variety of tangible benefits, such as spectrum efficiency \cite{zhang2021overview}, less expensive modem by unifying hardware and software resources across \ac{SnC} tasks \cite{9830717}. These advantages bring a significant number of difficulties and research issues that must be resolved, such as the trade-offs between high data rates and high-resolution sensing. One of the earliest appearances of \ac{ISAC} was in the 60s \cite{4337601}, where \ac{SnC} tasks share the same power, spectrum and resources. In particular, pulse interval modulation was used for radar pulses and communication was embedded in those pulses. Owing to advances in waveform design, beamforming, and \ac{SDRs}, \ac{ISAC} has become a reality.\\
\indent \ac{ISAC} can be categorized into three categories, from a design perspective. A \textit{joint design} approach aims at developing new waveforms intended for \ac{SnC}, while offering tradeoffs between different \ac{KPIs}, such as data rates, capacity, energy for communications and probability of detection and \ac{CRB} for radar sensing. For e.g., the work in \cite{bazzi2022integrated} proposes a joint design of \ac{SnC} waveforms with adjustable \ac{PAPR}, allowing trade-offs between radar performance, communication rate and \ac{PAPR}. Moreover, another design is \textit{radar-centric} \ac{ISAC}, where a radar waveform is considered and communication information is inserted onto the radar waveform \cite{7347464,9093221,7746569}. Finally, \textit{communication-centric} \ac{ISAC} takes a communication waveform, for example, \ac{OFDM}, and estimates sensing parameters based on channel information probed in that waveform. Here sensing is an "add-on". 
\ac{DFRC} is a design technique that integrates \ac{RnC} functions using a common transmit/receive aperture \cite{8828023}. For radar, there are two common settings for radar: mono-static, where transmit and receive units are co-located and bi-static, where the units are separated. 
In this paper, we propose a new scheme named \ac{HRF} that considers both mono-static and bi-static settings, where the \ac{BS} acts as a mono-static radar and users act as distributed bi-static radars in the \ac{UL} following a \ac{FDD} protocol.


\vspace{-0.1cm}
\subsection{Literature Review}
\label{subsec:literature-review}
To unbolt the potential of sensing capabilities in \ac{ISAC} systems, various lines of research work have been conducted in the literature. 
In particular, the paper in \cite{9468975} exploits \ac{OFDM} based waveforms in the \ac{DL} only, from a mono-static radar perspective, and performs delay and Doppler estimation via periodogram criterion, which is only optimal for a single target. 
In addition, \cite{9860521} considers a bi-static setting, in which \ac{SnC} tasks are performed in orthogonal sets of sub-carriers, where the sub-carriers allocated to sensing are distributed among antennas in an interleaved fashion. 
Besides, the paper in \cite{9937163} proposes an \ac{ISAC} multi-user \ac{IRS}-aided scheme, capable of carrying out \ac{UL} data communications, while simultaneously performing target localization. 
Nevertheless, \cite{9724260} investigates multi-beam systems for \ac{ISAC} for 5G-\ac{NR} using \ac{OFDM} waveforms. The \ac{BS} in this study is a mono-static one, and joint range and \ac{AoA} estimation are modeled as sensing parameters, which further assists the beam-scanning procedure. However, the framework utilizes only the \ac{DL} information, and therefore the \ac{UL} information available from different users is neglected. Meanwhile, \cite{10018908,liu2018toward} consider the joint design of beamforming for \ac{SnC} purposes, which is well-suited for beam-scanning applications. These frameworks also rely on \ac{DL}-only communication information for sensing purposes. Furthermore, from a physical-layer security \ac{ISAC} perspective, the work in \cite{su2022sensing} provides sensing estimation based on the eavesdropper location to improve the secrecy rate of the system. In short, the framework utilizes \ac{DL}-only information for the radar echo. To the best of our knowledge, there does not exist any \ac{ISAC} scheme in the literature that functions on a hybrid monostatic and bistatic mode and fuses \ac{UL} and \ac{DL} information to efficiently estimate target sensing parameters.
The work in \cite{9805471} focuses on a massive \ac{MIMO} \ac{ISAC} setting where the \ac{SnC} channels reveal joint sparsity.
It is worth highlighting that, in contrast to the approach discussed in \cite{9805471}, our methodology does not rely on a massive \ac{MIMO} setting, nor does it require a large number of antennas.
In addition, we do operate under the assumption of a \ac{FDD} communication protocol, where the \ac{DFRC} \ac{BS} further fuses channel estimates, hence, maximizing their utility for sensing purposes.
From an estimation perspective, and in order to capture the joint sparse burst structure of the scattering components for sensing, the work in \cite{9805471} employs a Turbo sparse bayesian learning algorithm, which shows promising results.
However, it was assumed that the discrete \ac{AoA} grid points for the dictionary (design) matrix should be set equal to the number of \ac{BS} antennas to provide a good trade-off between \ac{AoA} estimation and complexity. 
Consequently, achieving good performance necessitates a large number of antennas, presenting an advantageous prospect for massive \ac{MIMO} applications.
Additionally, the work in \cite{9898900} proposes a compressed sensing for sensing purposes, in an \ac{ISAC} massive \ac{MIMO} setting.
\vspace{-0.1cm}
\subsection{Contributions}
\label{subsec:contributions}
In this paper, we propose a \ac{HRF} system model, that combines both monostatic and bistatic modes, where the \ac{DFRC} \ac{BS} exploits the echos due to the \ac{DL} communication signals, in conjunction with \ac{UL} signals from a set of users associated with the \ac{DFRC} \ac{BS}. Each communication user transmits \ac{OFDM} waveforms on its own resource band, where the \ac{DFRC} \ac{BS} then fuses the information from different resources to estimate \ac{AoAs} of the targets in the scene.
Since we propose to estimate sensing parameters of targets without evaluating communication performance, the considered system falls under the communication-assisted sensing category. In other terms, our main interest lies in the estimation quality of target \ac{AoAs} with the aid of a well-established communication infrastructure. To this end, we outline our contributions as
\begin{itemize}
	\item \textbf{Hybrid Radar Model.} We coin the term \textit{hybrid} because of the diverse and distributed radar functionality considered in this paper, i.e. the \ac{DL} echo is associated with a mono-static radar setting, as the \ac{DFRC}'s transmit and receive units are assumed to be colocated. Meanwhile, a (user,\ac{DFRC}) couple is seen as a bi-static radar, as the user is seen as a radar transmit unit in the \ac{UL}, transmitting via an \ac{FDD} protocol through \ac{OFDM} communication signals, and the \ac{DFRC} receive unit is the radar receiver. The \ac{DFRC}'s sophisticated task is to fuse the hybrid data at its disposal to estimate target locations. \textit{To the best of our knowledge, the estimation model that arises due to the hybrid nature appears in the literature for the first time.}
	\item \textbf{Fused Maximum Likelihood Criterion.} We derive the \ac{ML} criterion of the hybrid radar model, and highlight our interpretations, such as its resemblance with the classical \ac{ML} criterion and the additional terms due to the participation of the communication users in the \ac{UL}.
	\item \textbf{Fused and Efficient Estimators.} In light of the hybrid and dual nature of the available data, we propose two computationally attractive estimators. The first one explicitly tackles the \ac{ML} criterion through an iterative optimization procedure that alternates between the target and user spaces. The second is an iterative subspace estimator: inspired by the \ac{MUSIC} principle, we formulate novel optimization problems, which lead to a new iterative estimator that carefully projects and fuses target and user location estimates.
	\item \textbf{Extensive simulation results.} Various performance metrics evaluating sensing performance via \ac{OFDM} communication waveforms are studied through simulations, demonstrating the effectiveness of utilizing fused data for the hybrid radar model. Numerical results show performance gains in terms of \ac{MSE} of the estimated \ac{AoAs} for the targets, especially when more communication users participate in the \ac{UL}. 
\end{itemize}
 
%
%

\textbf{Notation}: Upper-case and lower-case boldface letters denote matrices and vectors, respectively. $(.)^T$, $(.)^*$ and $(.)^H$ represent the transpose, the conjugate and the transpose-conjugate operators. The empty set is $\emptyset$. The cardinality of $\mathcal{C}$ is $\vert \mathcal{C} \vert $. The set of all complex-valued $N \times M$ matrices is $\mathbb{C}^{N \times M}$. The determinant is $\det()$. The trace is $\trace()$. The $N \times N $ identity matrix is $\pmb{I}_N$. The projector operator onto the space spanned by columns of $\pmb{A} \in \mathbb{C}^{N \times M}$ is $\pmb{P}_{\pmb{A}}$ and the corresponding orthogonal projector is $\pmb{P}_{\pmb{A}}^{\perp} \triangleq \pmb{I}_N - \pmb{P}_{\pmb{A}}$. We index the $(i,j)^{th}$ element of $\pmb{A}$ as $\pmb{A}(i,j)$. The $k^{th}$ column of $\pmb{A}$ is indexed as $\pmb{A}(:,k)$. The $\Span$ of $\pmb{A}$, denoted $\Span(\pmb{A})$ returns the column space of $\pmb{A}$. The Moore-Penrose pseudo-inverse of $\pmb{A}$ is $\pmb{A}^+$.

\section{System Model}
\label{sec:system-model}
\begin{figure}[!t]
\centering
\includegraphics[width=3in]{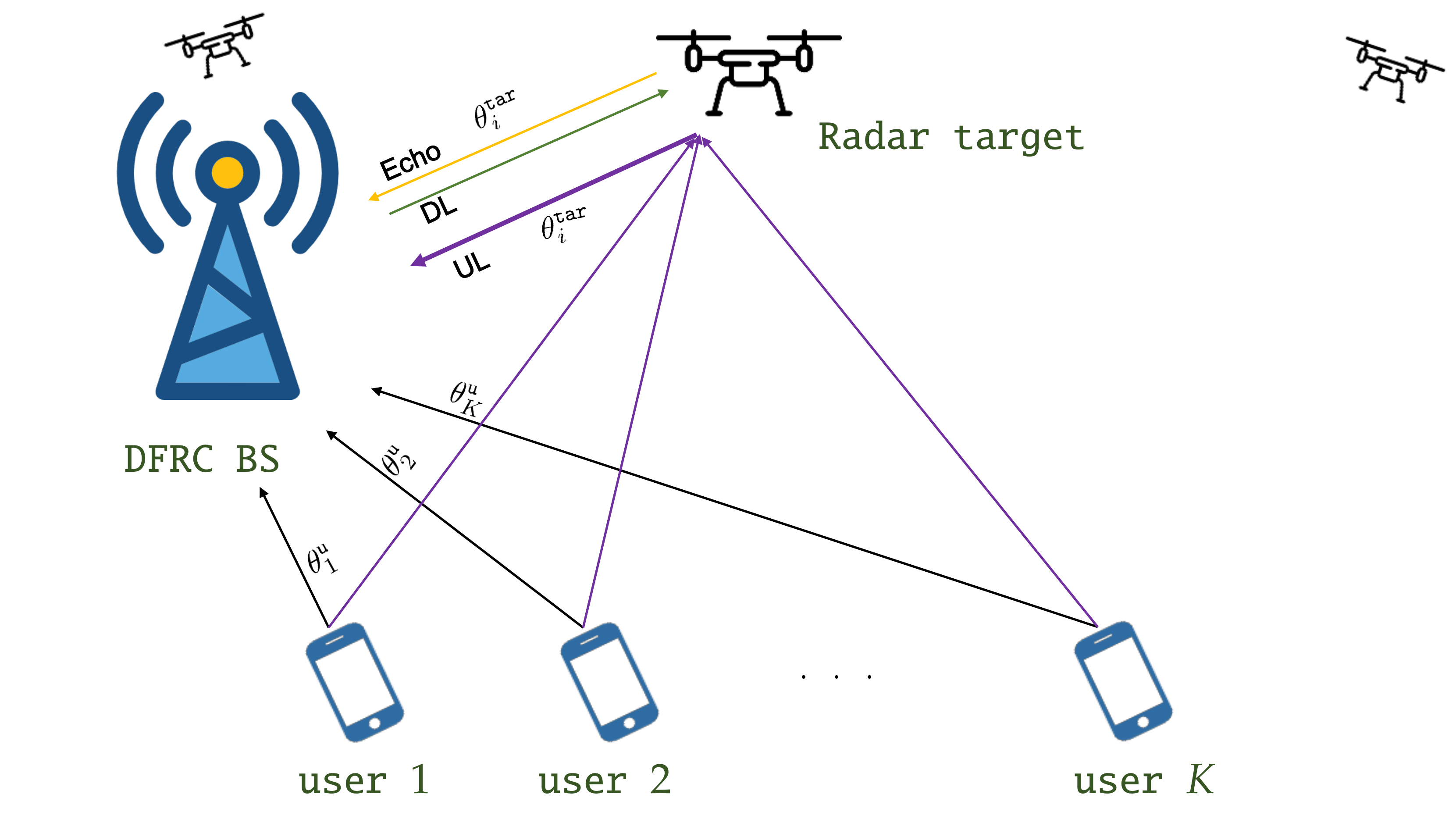}
\caption{The \ac{HRF} system model comprised of a \ac{DFRC} \ac{BS} in mono-static mode listening to its \ac{DL} echo, with $K$ communication \ac{UL} users and a radar target. We only depict one target due to illustration purposes.}
\label{fig_1}
\end{figure}

Consider a \ac{DFRC} \ac{BS} transmitting \ac{OFDM} symbols over its allocated band.
The \ac{DFRC} \ac{BS} is equipped with $N_t$ transmit antennas aiding in \ac{SnC} directional beamforming.
Furthermore, assume that $q$ targets are present in the scene, where the $i^{th}$ target is located at $\theta_i^{\tt{tar}}$ from the \ac{DFRC} \ac{BS}. We denote this set as $\pmb{\Theta}^{\tt{tar}} = \begin{bmatrix}
	\theta_1^{\tt{tar}} & \ldots & \theta_q^{\tt{tar}}
\end{bmatrix}$.In the \ac{DL}, the $\ell^{th}$ transmitted \ac{OFDM} signal can be expressed as 
\begin{equation}
	\label{eq:tx-OFDM-DL}
		\pmb{s}_{0,\ell}(t) = \Big( \sum\nolimits_{m \in \mathcal{C}_0 }b_{m,0}^{(\ell)} \exp(j 2 \pi m \Delta_f t )\Pi(t- \ell T) \Big) \pmb{f},
\end{equation}
where $\pmb{f} \in \mathcal{C}^{N_t \times 1}$ is the transmit beamforming vector.
Furthermore, $\mathcal{C}_0$ is the set of subcarriers corresponding to the allocated band in the \ac{DL}. Furthermore, $b_{m,0}^{(\ell)}$ is the symbol on the $m^{th}$ \ac{OFDM} subcarrier within the $\ell^{th}$ \ac{OFDM} symbol and $\Delta_f$ is the subcarrier spacing. 
Also, the number of \ac{OFDM} symbols in the \ac{DL} is denoted as $L_0$.
Moreover, the \ac{OFDM} symbol duration is given $T = \frac{1}{\Delta_f}$. Also, $\Pi(t)$ is a certain windowing function. The overall \ac{OFDM} symbol duration is denoted as $T_o \triangleq T + T_{\text{CP}}$, where $T_{\text{CP}}$ is the cyclic prefix duration. Assuming the ideal rectangular function, the window function should satisfy $\Pi(t) = 1$, if $t \in [-T_{\text{CP}},T]$, and $0$, else. Here, $T_{\text{CP}}$ should be greater than the maximum of propagation delays in order to guarantee a cyclic convolution with the channel. The $\ell^{th}$ transmitted \ac{OFDM} symbol from the single-antenna $k^{th}$ user in the \ac{UL} is given by
\begin{equation}
	\label{eq:tx-OFDM-UL}
	s_{k,\ell}(t) = 
	\sum\nolimits_{m \in \mathcal{C}_k }
	b_{m,k}^{(\ell)} \exp(j 2 \pi m \Delta_f t )
	\Pi(t- \ell T),
\end{equation}
where $\mathcal{C}_k$ is the set of subcarriers corresponding to the allocated band for the $k^{th}$ user in the \ac{UL}. Moreover, $b_{m,k}^{(\ell)}$ is the symbol on the $m^{th}$ \ac{OFDM} subcarrier by the $k^{th}$ user within the $\ell^{th}$ \ac{OFDM} symbol. Also, the transmission consists of $L_k$ \ac{OFDM} symbols for the $k^{th}$ user. We have that $\mathcal{C}_0 \cup \mathcal{C}_1 \cup \ldots \cup \mathcal{C}_K = \lbrace 0,1, \ldots, M-1 \rbrace$, where $M$ is the total number of subcarriers occupied by the entire system, i.e. the total occupied bandwidth is $M\Delta_f$. We also assume that the subcarrier sets are non-overlapping, i.e. $\mathcal{C}_i \cap \mathcal{C}_j = \emptyset$ for all $i \neq j$.
Assuming a colocated mono-static \ac{MIMO} radar setting, the time-domain baseband channel at a given time $t$ and at the $n^{th}$ receive antenna can be represented as follows
\begin{equation}
	\label{eq:channel-DL-echo}
	\pmb{h}_n^{\tt{echo}}(t)
	=
	\sum\nolimits_{i \in \Phi_0}
	g_{i}^{\tt{tar}}
	a_n(\theta_i^{\tt{tar}})
	\pmb{a}_t^T(\theta_i^{\tt{tar}})
	\delta(t- 2 \tau_i^{\tt{tar}})
	e^{j2\pi f_{D,i} t},
\end{equation}where $\Phi_0$ denotes the set of target indices explored by the \ac{DFRC} \ac{BS}. The \ac{ToA} between the \ac{DFRC} \ac{BS} and the $i^{th}$ target is denoted as $\tau_i^{\tt{tar}}$.
The frequency $f_{D,i}$ represents the Doppler frequency of the $i^{th}$ target, which arises due to the movement of the target.
The vector $\pmb{a}_t(\theta) \in \mathbb{C}^{N_t \times 1}$ is the transmit steering vector pointing towards \ac{AoD} $\theta$. Notice that due to the mono-static configuration, the \ac{AoA} and \ac{AoD} of a certain target are the same.The general expression of the steering response of the $n^{th}$ antenna due to a signal arriving at angle $\theta$, $a_n(\theta)$, is generally dependent on the frequency. Its most general form follows
\begin{equation}
	a_n(\theta,f)
	=
	\exp
	\Big(
	j 2 \pi 
	\frac{f}{c}
	(x_n \sin(\theta) + y_n \cos(\theta))
	\Big),
\end{equation}
where $(x_n,y_n)$ represent the cartesian coordinates of antenna $n$ in the $(x,y)-$ plane and $\theta$ is the \ac{AoA}. 
In this paper, we follow a \textit{narrowband assumption}, i.e. the bandwidth occupied by the \ac{FDD} protocol, $B = \sum_{k=0}^K \vert \mathcal{C}_k \vert \Delta_f$ is much smaller than the carrier frequency $f_c$, i.e. $\frac{B}{2f_c} \ll 1$.
Hence, $a_n(\theta,f) \simeq  a_n(\theta) = \exp
	\big(
	j 2 \pi 
	\frac{f_c}{c}
	(x_n \sin(\theta) + y_n \cos(\theta))
	\big)$.
A similar discussion can be done on the transmit steering vector $\pmb{a}_t(\theta)$.
 The term $g_i^{\tt{tar}}$ is the two-way propagation complex gain of the $i^{th}$ target which also contains an initial phase offset. Its magnitude is given by the two-way path-loss equation as 
\begin{equation}
	\label{eq:PL-echo}
	\vert g_i^{\tt{tar}} \vert
	=
	\sqrt{
	\frac{\lambda^2 \sigma^{{\tt{RCS}}}_i \beta_{\tt{Tx}}(\theta_i^{\tt{tar}})
	\beta_{\tt{Rx}}(\theta_i^{\tt{tar}})}{(4\pi)^3 (d_i^{\tt{tar}})^4}},
\end{equation}
where $\sigma^{{\tt{RCS}}}_i$ is the \ac{RCS} of the $i^{th}$ target and $d_i^{\tt{tar}}$ is the Euclidean distance between the $i^{th}$ target and the \ac{DFRC} \ac{BS}. In fact, $\tau_i^{\tt{tar}} = \frac{d_i^{\tt{tar}}}{c}$, where $c$ is the speed of light. The terms $\beta_{\tt{Tx}}(\theta)$ and $\beta_{\tt{Rx}}(\theta)$ represent the gain of the transmit and receive antenna elements at look-direction $\theta$, respectively. The wavelength is $\lambda$.

On the other hand, the \ac{UL} channel between the $k^{th}$ communication user and the \ac{DFRC} \ac{BS} is characterized as follows
\begin{equation}
	\label{eq:channel-UL}
\begin{split}
	h_{n,k}^{\tt{UL}}(t)
	&=
	g_{0,k}^{\tt{u}}
	a_n(\theta_k^{\tt{u}})
	\delta(t- \tau_{k}^{\tt{u}})\\ &
	+
	\sum\nolimits_{i \in \Phi_k}
	g_{i,k}^{\tt{tar}}
	a_n(\theta_i^{\tt{tar}})
	\delta(t- \phi_{k,i})
	e^{j2\pi f_{D,i} t},
\end{split}
\end{equation}
where the first term is the \ac{LoS} channel between the $k^{th}$ user and the \ac{DFRC} \ac{BS}. 
The second term constitutes the two-way channel between the $k^{th}$ user towards the set of targets $\Phi_k$, where $\Phi_k$ denotes the set of target indices explored by the $k^{th}$ user.
The indices within $\Phi_k$ are denoted as $\Phi_k = \begin{bmatrix}
	u_1^{(k)}, & \ldots ,&u^{(k)}_{\vert \Phi_k \vert} 
\end{bmatrix}$, for all $k = 0 \ldots K$. 
Furthermore, we assume that the union of indices defined by the set of targets explored by the \ac{DFRC} \ac{BS} and all users cover all the $q$ targets, namely $\bigcup_{k} \Phi_k = \lbrace 1 \ldots q \rbrace$.
In general, $\vert \Phi_k \vert  \leq \vert \Phi_0\vert $ for $k = 1,2,\cdots, K$, and $\Phi_i\neq \Phi_j$ for $i\neq j$.The \ac{AoA} between the $k^{th}$ user and \ac{BS} is $\theta_k^{\tt{u}}$. The path gains are given as
 \begin{align}
 	\label{eq:PL-LoS}
 	\vert g_{0,k}^{\tt{u}} \vert
 	&=
 	\sqrt{
 	\frac{\lambda^2 \beta_{\tt{Tx}}^{\tt{u}}(\theta_k^{\tt{u}}) \beta_{\tt{Rx}}(\theta_k^{\tt{u}})}{(4\pi)^2 (d_k^{\tt{u}})^2}}, \\
 	\label{eq:PL-UL}
  	\vert g_{i,k}^{\tt{tar}} \vert
 	&
 	=
 	\sqrt{
 	\frac{\lambda^2 \bar{\sigma}^{{\tt{RCS}}}_{i,k}\beta_{\tt{Tx}}^{\tt{u}}(\psi_{k,i}) \beta_{\tt{Rx}}(\theta_i^{\tt{tar}})}{(4\pi)^3 (d_{k,i})^2 (d_i^{\tt{tar}})^2}},
 \end{align}
where $\beta_{\tt{Tx}}^{\tt{u}}(\theta)$ is the gain of the user's antenna towards direction $\theta$, $\psi_{k,i}$ and $d_{k,i}$ represent the angle and distance between the $k^{th}$ user and the $i^{th}$ target, respectively. 
In addition, $\bar{\sigma}^{{\tt{RCS}}}_{i,k}$ is the \ac{RCS} of the $i^{th}$ target due to the $k^{th}$ communication user.
The term $d_k^{\tt{u}}$ is the distance between the $k^{th}$ user and the \ac{DFRC} \ac{BS}. In addition, we denote the \ac{ToA} between the $k^{th}$ user and the \ac{DFRC} \ac{BS} as $\tau_k^{\tt{u}} = \frac{d_k^{\tt{u}}}{c}$. Also, the two-way delay $\phi_{k,i}$ is the delay from $k^{th}$ user towards $i^{th}$ target then towards the \ac{DFRC} \ac{BS}, i.e. $\phi_{k,i} = \frac{d_{k,i} + d_i^{\tt{tar}}}{c} = \tau_{k,i}+\tau^{\tt{tar}}_i$, and $\tau_{k,i}$ is the one-way delay between the $k^{th}$ user and $i^{th}$ target.

The $\ell^{th}$ \ac{OFDM} symbol at the $n^{th}$ receive antenna of the \ac{BS} after performing convolution and downconversion can be written as
\begin{equation}
	\label{eq:rx-OFDM}
	r_{n,\ell}(t) 
	=
	r_{n,\ell}^{\tt{echo}}(t)
	+
	\sum\nolimits_{k=1}^K
	r_{n,\ell,k}^{\tt{UL}}(t)
	+
	w_{n,\ell}(t),
\end{equation}
where $r_{n,\ell}^{\tt{echo}}(t)$ is the received echo signal due to the presence of the radar targets, i.e.
\begin{equation}
	\label{eq:rx-OFDM-echo}
		r_{n,\ell}^{\tt{echo}}(t) 
	=
	\sum\nolimits_{i \in \Phi_0} 
	\gamma_{i,\ell}^{\tt{tar}} 
	(\pmb{a}_t^T(\theta_i^{\tt{tar}}) \pmb{f})
	a_n(\theta_i^{\tt{tar}})
	s_{0,\ell}(t - 2\tau_i^{\tt{tar}}).
\end{equation}
Furthermore, $r_{n,\ell,k}^{\tt{UL}}(t)$ is the received signal due to the channel between the $k^{th}$ user and the \ac{DFRC} \ac{BS}, taking into account the direct \ac{LoS} and bounces due to the presence of targets, i.e.
\begin{equation}
	\label{eq:rx-OFDM-UL}
\begin{split}
	r_{n,\ell,k}^{\tt{UL}}(t) 
	&=
	\gamma_{0,\ell,k}^{\tt{u}}
	a_n(\theta_k^{\tt{u}})
	s_{k,\ell}(t-\tau_k^{\tt{u}}) \\
	&+
	\sum\nolimits_{i\in\Phi_k} 
	\gamma_{i,\ell,k}^{\tt{tar}} 
	a_n(\theta_i^{\tt{tar}})
	s_{k,\ell}(t - \phi_{k,i}).
\end{split}
\end{equation}
In the above, we have approximated the varying phases due to Doppler frequency over an \ac{OFDM} block symbol as constant \cite{zhang2021overview}, therefore, 
$\gamma_{i,\ell}^{\tt{tar}} = g_{i}^{\tt{tar}} e^{j 2 \pi \ell T f_{D,i}}$.
Likewise, $\gamma_{i,\ell,k}^{\tt{tar}} = g_{i,k}^{\tt{tar}} e^{j 2 \pi \ell T f_{D,i}}$ and $\gamma_{0,\ell,k}^{\tt{u}} = g_{0,k}^{\tt{u}} 
$.
The background noise in \eqref{eq:rx-OFDM} is denoted as $w_{n,\ell}(t)$, which is assumed to be \ac{i.i.d} \ac{AWGN}. Now, plugging equations \eqref{eq:tx-OFDM-DL} and \eqref{eq:tx-OFDM-UL} in equations \eqref{eq:rx-OFDM-echo} and \eqref{eq:rx-OFDM-UL}, and sampling at regular intervals of $p \triangleq p \frac{T}{M}$, we can express equation \eqref{eq:rx-OFDM} as follows

\begin{equation}
\label{eq:sample-rx-OFDM}
	\begin{split}
		r_{n,\ell}[p] 
	&=
	\sum\limits_{\substack{i\in\Phi_0 \\ m \in \mathcal{C}_0 }} 
	b_{m,0}^{(\ell)}
	e^{j 2 \pi  \frac{m  p}{M} }
	c_m(2\tau_i^{\tt{tar}})
	\gamma_{i,\ell}^{\tt{tar}} 
	(\pmb{a}_t^T(\theta_i^{\tt{tar}}) \pmb{f})
	a_n(\theta_i^{\tt{tar}}) \\
	&+
	\sum\limits_{\substack{k=1 \ldots K \\ m \in \mathcal{C}_k}}
	b_{m,k}^{(\ell)}
	e^{j 2 \pi  \frac{m  p}{M} }
	c_m(\tau_k^{\tt{u}})
	\gamma_{0,\ell,k}^{\tt{u}}
	a_n(\theta_k^{\tt{u}})  \\
	&+
	\sum\limits_{\substack{k=1 \ldots K \\ m \in \mathcal{C}_k \\ i\in\Phi_k}}
	b_{m,k}^{(\ell)}
	e^{j 2 \pi  \frac{m  p}{M} }
	c_m(\phi_{k,i})
	\gamma_{i,\ell,k}^{\tt{tar}} 
	a_n(\theta_i^{\tt{tar}}) +w_{n,\ell}[p],
	\end{split}
\end{equation}
where $c_m(\tau) = e^{-j 2 \pi m \Delta_f \tau}$. Gathering $M$ time samples of $r_{n,\ell}[p]$ and performing a \ac{DFT}, we have 
\begin{equation}
\label{eq:DFT-formula}
	R_{n,\ell,m}
	=
	\frac{1}{M}
	\sum\nolimits_{p=0}^{M-1}
	r_{n,\ell}[p] 
	e^{-j 2 \pi  \frac{m  p}{M}}.
\end{equation}
%
Due to the above \ac{DFT}, and thanks to the orthogonality in frequency domain due to \ac{FDD} operation, i.e. $\mathcal{C}_i \cap \mathcal{C}_j = \emptyset$ for all $i \neq j$, we have
\begin{equation}
\label{eq:equalization}
		Y_{n,\ell,m,k}
	=
	R_{n,\ell,m} \Big\vert_{m \in \mathcal{C}_k}, \quad 
	k = 0 \ldots K
\end{equation}
where
\begin{equation}
\label{eq:DL-after-equalization}
	\begin{split}
	Y_{n,\ell,m,0}
	&=
	\sum\nolimits_{i \in \Phi_0} c_m(2\tau_i^{\tt{tar}})
	\gamma_{i,\ell}^{\tt{tar}} 
	(\pmb{a}_t^T(\theta_i^{\tt{tar}}) \pmb{f})
	a_n(\theta_i^{\tt{tar}})\\
	&+
	W_{n,\ell,m,0},	
\end{split}
\end{equation}
for $m \in \mathcal{C}_0$, and
\begin{equation}
\label{eq:UL-after-equalization}
	\begin{split}
	Y_{n,\ell,m,k}
	&=		
	c_m(\tau_k^{\tt{u}})
	\gamma_{0,\ell,k}^{\tt{u}}
	a_n(\theta_k^{\tt{u}}) \\
	&+
	\sum\nolimits_{i \in \Phi_k} 
	c_m(\phi_{k,i})
	\gamma_{i,\ell,k}^{\tt{tar}} 
	a_n(\theta_i^{\tt{tar}}) +
	W_{n,\ell,m,k},
\end{split}
\end{equation}
for $m \in \mathcal{C}_k$ for all $k \geq 1$. Without loss of generality, we have assumed that $b_{m,k}^{(\ell)} = 1$, as those are known pilot symbols.Equations \eqref{eq:DL-after-equalization} and \eqref{eq:UL-after-equalization} are arranged, via matrix form, namely
\begin{align}
\label{eq:matrix-form-UL}
	\pmb{Y}_0
	&= 
	\pmb{A}(\pmb{\Theta}^{\tt{tar}}_0)
	\pmb{X}_0
	+
	\pmb{W}_0, \\
\label{eq:matrix-form-DL}
	\pmb{Y}_k
	&= 
	\begin{bmatrix}
		\pmb{a}(\theta_k^{\tt{u}})		&
		\pmb{A}(\pmb{\Theta}^{\tt{tar}}_k)
	\end{bmatrix}
	\pmb{X}_k + \pmb{W}_k, \quad k = 1 \ldots K.
\end{align}
For \ac{UL} communications, the preamble $s_{k,\ell}(t)$ is usually precedes a payload. The payload can be decoded by using the channel estimate per user, i.e. $Y_{n,\ell,m,k}$. 
Let $\mathcal{C}_k = \lbrace p_1, p_2, \ldots , p_{\vert \mathcal{C}_k\vert} \rbrace$ be the subcarrier indices occupied by the $k^{th}$ user in the \ac{UL}. To this extent, we define $\pmb{X}_k$ as   
\begin{equation}
\label{eq:Xk-matrix-form}
	\pmb{X}_k
	=
	\begin{bmatrix}
		\pmb{x}_{1,p_1}^{(k)}
		,
		\ldots
		,
		\pmb{x}_{L_k,p_1}^{(k)}
		,
		\ldots
		,
		\pmb{x}_{1,p_{\vert \mathcal{C}_k\vert}}^{(k)}
		,
		\ldots
		,
		\pmb{x}_{L_k,p_{\vert \mathcal{C}_k\vert}}^{(k)}
	\end{bmatrix},
\end{equation}
where $\pmb{x}_{\ell,m}^{(k)}$ is a function of the gains, delays, as well as Doppler shifts.
Generally speaking, the vector $\pmb{x}_{\ell,m}^{(k)}$ can also be modeled to contain synchronization parameters caused by misalignment of oscillator clocks between the \ac{DFRC} \ac{BS} and the  \ac{UL} users, giving rise to time offsets and frequency offsets. Hence, these parameters may not have severe impact on the \ac{AoA} estimation performance. In this paper, we treat variables $\pmb{x}_{\ell,m}^{(k)}$ as nuisance parameters with arbitrary structure, allowing for its estimation and compensation at a following stage.Notice that $\pmb{X}_k \in \mathbb{C}^{(\vert \Phi_k \vert +1) \times \bar{L}_k}, \forall k = 1 \ldots K$ and $\bar{L}_k = L_k \vert \mathcal{C}_k \vert $ is the overall number of snapshots seen by the \ac{DFRC} \ac{BS} from the $k^{th}$ user in the \ac{UL}. Likewise, $\pmb{X}_0 \in \mathbb{C}^{\vert \Phi_0 \vert \times \bar{L}_0}$ and $\bar{L}_0$ is the total number of snapshots seen through the transmitted \ac{DL} \ac{OFDM} frame. Furthermore, the manifold matrix $\pmb{A}(\pmb{\Theta}^{\tt{tar}}_k)$ is
\begin{equation}
	\label{eq:kth-manifold}
		\pmb{A}(\pmb{\Theta}^{\tt{tar}}_k)
	=
	\begin{bmatrix}
		\pmb{a}(\theta_{u^{(k)}_1}^{\tt{tar}}) & \ldots &
		\pmb{a}(\theta_{u^{(k)}_{\vert \Phi_k \vert}}^{\tt{tar}})
	\end{bmatrix}
	\in \mathbb{C}^{N \times \vert \Phi_k \vert},
\end{equation}
and the steering vector is given by 
\begin{equation}
	\pmb{a}(\theta)
	=
	\begin{bmatrix}
		\pmb{a}_1(\theta) & \ldots & \pmb{a}_N(\theta) 
	\end{bmatrix}^T \in \mathbb{C}^{N \times 1}.
\end{equation}
Moreover, the entries of $\pmb{W}_k$ contain $W_{n,\ell,m,k}$ for $m \in \mathcal{C}_k$. In this paper, since we are interested in target \ac{AoA} estimation, we shall treat the entries of $\pmb{X}_0,\pmb{X}_1 \ldots \pmb{X}_K$ as nuisance parameters, as the estimation of complex channel gains, \ac{ToA} and Doppler parameters is not the main focus herein.

Furthermore, observing equations \eqref{eq:matrix-form-UL} and \eqref{eq:matrix-form-DL}, one can directly relate each equation to the classical \ac{AoA} estimation problem. However, we would like to stress the fundamental difference between the \ac{AoA} problem at hand and the classical one. Even though each of the $K+1$ matrix equations in \eqref{eq:matrix-form-UL} and \eqref{eq:matrix-form-DL} are classical \ac{AoA}-type equations, the information contained over all equations is not. More specifically, each user over its occupied bandwidth, defined by $\mathcal{C}_k$, contains the manifold of interest $\pmb{A}(\pmb{\Theta}^{\tt{tar}}_k)$ appended by an unknown channel $\pmb{a}(\theta_k^{\tt{u}})$, linking the $k^{th}$ user towards the \ac{DFRC} \ac{BS}. 

Our system model is illustrated in Fig. \ref{fig_1}. We can see that the \ac{DFRC} \ac{BS} broadcasts the \ac{DL} data and listens to the received echo at its allocated band. Meanwhile, the users each transmit \ac{OFDM} symbols within their own band in the \ac{UL}. Each communication contributes in two components:(i) a \ac{LoS} path between itself and the \ac{DFRC} \ac{BS}, which is assumed to be unknown and (ii) the bounce from the target towards the \ac{DFRC} \ac{BS} within the users allocated band. The \ac{DFRC} \ac{BS} would then fuse the information available at all the bands, in order to estimate the unknown \ac{AoAs}.

Before we introduce the estimation criteria and methods, we would like to take a moment and observe the joint model appearing in equations \eqref{eq:matrix-form-UL} and \eqref{eq:matrix-form-DL}. The joint consideration of the $K+1$ models is the first time to be addressed, from an estimation perspective. Indeed, if we regard each of the $K+1$ models separately, then we coincide with a classical \ac{AoA} estimation problem. 
However, the fused processing of all the $K+1$ models can be leveraged to our advantage, due to the presence of a \textit{common manifold} being shared across all the bands, $\pmb{A}(\pmb{\Theta}^{\tt{tar}}_{\cap})$, where $\pmb{\Theta}^{\tt{tar}}_{\cap} =  \pmb{\Theta}_0^{\tt{tar}} \cap \pmb{\Theta}_1^{\tt{tar}} \cap \ldots \cap \pmb{\Theta}_K^{\tt{tar}}$.
 To this end, we are ready to address our problem: by fusing the observations $\lbrace \pmb{Y}_0,\pmb{Y}_1 \ldots \pmb{Y}_K \rbrace$ as defined in \eqref{eq:matrix-form-UL} and \eqref{eq:matrix-form-DL}, we would like to estimate $\pmb{\Theta}^{\tt{tar}}$, as well as $\pmb{\Theta}^{\tt{u}}$.

\vspace{-0.1cm}
\section{Fused Maximum Likelihood}
\label{sec:fused-maximum-likelihood}
In this section, we derive the deterministic \ac{ML} for estimating $\pmb{\Theta}^{\tt{tar}}$ and $\pmb{\Theta}^{\tt{u}}$. The deterministic \ac{ML} regards the sample functions as unknown deterministic sequences, rather than random processes. As shall be clarified, we can interpret our \ac{ML} criterion as a \ac{FML} one. 
Denoting $\pmb{Y} = \begin{bmatrix}
	\pmb{Y}_0 & \ldots & \pmb{Y}_K
\end{bmatrix}$, we express the joint density function of the observed data per allocated bandwidth as
\begin{equation}
\label{eq:likelihood}
\begin{split}
	f(\pmb{Y})
	=&
	\prod_{k=0}^K
	\prod_{p=1}^{\bar{L}_k}
	\frac{1}{\pi \det(\sigma^2 \pmb{I})} \\
	&\times \exp 
	\Big(
	-\frac{1}{\sigma^2}
	\big\Vert 
	\pmb{Y}_k(:,p) 
	-
	\pmb{A}_k
	\pmb{X}_k(:,p)
	\big\Vert^2
	\Big),
\end{split}
\end{equation}
where, for notation's sake, we denote 
\begin{equation}
\label{eq:Ak-definition}
\pmb{A}_k
=
	\begin{cases}
		\pmb{A}(\pmb{\Theta}^{\tt{tar}}_0) &, k = 0 \\
		\begin{bmatrix}
		\pmb{a}(\theta_k^{\tt{u}})		&
		\pmb{A}(\pmb{\Theta}^{\tt{tar}}_k)
	\end{bmatrix}&, 1 \leq k \leq K.
\end{cases}
\end{equation}
Note that $f(\pmb{Y})$ is conditioned over $\pmb{X}_0\ldots\pmb{X}_K,\sigma^2,\pmb{\Theta}^{\tt{tar}},\pmb{\Theta}^{\tt{u}}$, and this dependency has been omitted for sake of compact notation.
Ignoring constant terms, the $\log$-likelihood function, $\mathcal{L} \triangleq \log f(\pmb{Y})$,  can be expressed as
\begin{equation}
\label{eq:log-likelihood}
\begin{split}
	\mathcal{L} \sim 
	-N\bar{L} \log \sigma^2
		-\frac{1}{\sigma^2}
	\sum\nolimits_{p,k}
	\big\Vert 
	\pmb{Y}_k(:,p) 
	-
	\pmb{A}_k
	\pmb{X}_k(:,p)
	\big\Vert^2,
\end{split}
\end{equation}
where $\bar{L} = \sum_{k=0}^K \bar{L}_k$. The \ac{ML} criterion is now given as 
\begin{equation}
\label{eq:maximum-likelihood}
\big\lbrace 
\widehat{\sigma}^2,
\widehat{\pmb{X}},
\widehat{\pmb{\Theta}}^{\tt{tar}},
\widehat{\pmb{\Theta}}^{\tt{u}} 
\big\rbrace
=
	\argmax\nolimits_{\lbrace \sigma^2,\pmb{X},\pmb{\Theta}^{\tt{tar}} , \pmb{\Theta}^{\tt{u}} \rbrace }
	\mathcal{L}
\end{equation}
where $\pmb{X} = \begin{bmatrix}
	\pmb{X}_0 & \pmb{X}_1 & \ldots & \pmb{X}_K
\end{bmatrix}$. To proceed, we first carry out the maximization process over $\sigma^2$ by fixing $\pmb{X},\pmb{\Theta}^{\tt{tar}} , \pmb{\Theta}^{\tt{u}}$. This gives us 
\begin{equation}
\label{eq:ML-sigma2}
	\widehat{\sigma}^2
	=
	\frac{1}{N\bar{L}}
	\sum\nolimits_{p,k}
	\big\Vert 
	\pmb{Y}_k(:,p) 
	-
	\pmb{A}_k
	\pmb{X}_k(:,p)
	\big\Vert^2.
\end{equation}
Replacing \eqref{eq:ML-sigma2} back into the $\log$-likelihood function and ignoring constant terms again, we get
\begin{equation}
\label{eq:maximum-likelihood-2}
	-N\bar{L} \log 
	\Big(
\frac{1}{N\bar{L}}
	\sum\nolimits_{k=0}^K
	\sum\nolimits_{p=1}^{\bar{L}_k}
	\big\Vert 
	\pmb{Y}_k(:,p) 
	-
	\pmb{A}_k
	\pmb{X}_k(:,p)
	\big\Vert^2
	\Big).
\end{equation} 
Using the monotonic property of the $\log$ function, the optimization criterion in \eqref{eq:maximum-likelihood-2} is now 
\begin{equation}
\label{eq:maximum-likelihood-3}
\big\lbrace 
\widehat{\pmb{X}},
\widehat{\pmb{\Theta}}^{\tt{tar}},
\widehat{\pmb{\Theta}}^{\tt{u}} 
\big\rbrace
=
	\argmin_{\lbrace \pmb{X},\pmb{\Theta}^{\tt{tar}} , \pmb{\Theta}^{\tt{u}} \rbrace }
	\sum\nolimits_{p,k}
	\big\Vert 
	\pmb{Y}_k(:,p) 
	-
	\pmb{A}_k
	\pmb{X}_k(:,p)
	\big\Vert^2.
\end{equation}
Now fixing $\pmb{\Theta}^{\tt{tar}} , \pmb{\Theta}^{\tt{u}} $, the optimization with respect of $\pmb{X}_k$ is the well known \ac{LS} over the $k^{th}$ user, i.e.
\begin{equation}
\label{eq:ML-X}
	\widehat{\pmb{X}}_k
	=
	\big(
	\pmb{A}_k^H
	\pmb{A}_k
	\big)^{-1}
	\pmb{A}_k^H
	\pmb{Y}_k.
\end{equation}
Note that the estimator of $\pmb{X}_k$ in \eqref{eq:ML-X} does not correspond to the exact \ac{ML}. Instead, it corresponds to a relaxed version of \ac{ML} by treating $\pmb{X}_k$ as an arbitrary matrix instead of considering its special structure as per equations \eqref{eq:DL-after-equalization} and \eqref{eq:UL-after-equalization}. Regarding $\pmb{X}_k$ as a nuissance parameter comes with some benefits: First, we bypass the need of estimating gain, delay and Doppler parameters related to each user and target. Second, this assumption simplifies the complexity of the resulting \ac{ML} estimator. Plugging \eqref{eq:ML-X} into \eqref{eq:maximum-likelihood-3}, we get
\begin{equation}
\label{eq:maximum-likelihood-4}
\big\lbrace 
\widehat{\pmb{\Theta}}^{\tt{tar}},
\widehat{\pmb{\Theta}}^{\tt{u}} 
\big\rbrace
=
	\argmin_{\lbrace \pmb{\Theta}^{\tt{tar}} , \pmb{\Theta}^{\tt{u}} \rbrace }
	\sum\nolimits_{p,k}
	\big\Vert 
	\pmb{Y}_k(:,p) 
	-
	\pmb{P}_k
	\pmb{Y}_k(:,p)
	\big\Vert^2,
\end{equation}	
where $\pmb{P}_k$ is the projector matrix projecting onto the column space of matrix $\pmb{A}_k$, namely
\begin{equation}
\label{eq:the-proj-matrix}
	\pmb{P}_k
	=
	\pmb{A}_k
	\big(
	\pmb{A}_k^H
	\pmb{A}_k
	\big)^{-1}
	\pmb{A}_k^H.
\end{equation}
The criterion in \eqref{eq:maximum-likelihood-4} can be replaced as
\begin{equation}
\label{eq:maximum-likelihood-5}
\big\lbrace 
\widehat{\pmb{\Theta}}^{\tt{tar}},
\widehat{\pmb{\Theta}}^{\tt{u}} 
\big\rbrace
=
	\argmax_{\lbrace \pmb{\Theta}^{\tt{tar}} , \pmb{\Theta}^{\tt{u}} \rbrace }
	\sum\nolimits_{p,k}
	\big\Vert 
	\pmb{P}_k
	\pmb{Y}_k(:,p)
	\big\Vert^2,
\end{equation}	
A more compact and convenient  form to express the above \ac{ML} cost is via the trace operator, i.e. 
\begin{equation}
\label{eq:maximum-likelihood-6}
\big\lbrace 
\widehat{\pmb{\Theta}}^{\tt{tar}},
\widehat{\pmb{\Theta}}^{\tt{u}} 
\big\rbrace
=
	\argmax_{\lbrace \pmb{\Theta}^{\tt{tar}} , \pmb{\Theta}^{\tt{u}} \rbrace }
	\sum\nolimits_{k=0}^K
	\trace 
	\Big(
	\pmb{P}_k
	\widehat{\pmb{R}}_k
	\Big),
\end{equation}	
where $\widehat{\pmb{R}}_k$ is the sample covariance matrix of $\pmb{Y}_k$,
\begin{equation}
\label{eq:sample-covariance-matrix}
	\widehat{\pmb{R}}_k = 
	\pmb{Y}_k
	\pmb{Y}_k^H.
\end{equation}
We coin the term \ac{FML}, as the \ac{ML} criterion naturally fuses information on the different bands specified by $\lbrace \mathcal{C}_k \rbrace_{k=0}^{K}$. It is worth noting that, although the different bands associated with users appear to be independent from an estimation viewpoint, this is not the case. Due to the fact that $\pmb{A}_0, \pmb{A}_1 \ldots \pmb{A}_K$ contain a common subspace $\pmb{A}(\pmb{\Theta}^{\tt{tar}}_{\cap})$, 
the samples over different bands can, to some extent, aid the estimation quality of $\widehat{\pmb{\Theta}}^{\tt{tar}}$, even though each band adds another unknown. In other words, the $k^{th}$ user introduces an unknown $\theta_k^{\tt{u}}$ within samples collected from that user. To highlight this even further, we invoke the block-wise formula of the projector matrix \cite{rao2008linear}, i.e.
\begin{equation}
\label{eq:projec-matr-decomp}
	\pmb{T}_{[\pmb{A} , \pmb{B}]}
	=
	\pmb{T}_{\pmb{A}}
	+
	\pmb{T}_{\pmb{B}_{\pmb{A}}}
	=
	\pmb{T}_{\pmb{A}}
	+
	\pmb{T}_{
	\pmb{T}_{\pmb{A}}^{\perp}
	\pmb{B}},
\end{equation}
where $\pmb{T}_{\pmb{A}}^{\perp} = \pmb{I}_N - \pmb{T} _{\pmb{A}}$.Now, we express $\pmb{P}_k$ in the \eqref{eq:maximum-likelihood-6} as 
\begin{equation}
\label{eq:projector-after-decomposition}
	\pmb{P}_k
	=
	\begin{cases}
		\pmb{P}_{\pmb{A}(\pmb{\Theta}^{\tt{tar}}_0)}, & k = 0 \\
		\pmb{P}_{\pmb{A}(\pmb{\Theta}^{\tt{tar}}_k)} 
		+
		\pmb{P}_{
		\pmb{P}_{\pmb{A}(\pmb{\Theta}^{\tt{tar}}_k)}^{\perp} \pmb{a}(\theta_k^{\tt{u}})} , & 1 \leq k \leq K
	\end{cases}
\end{equation}
Finally, plugging \eqref{eq:projector-after-decomposition} in the \ac{FML} cost in \eqref{eq:maximum-likelihood-6} and using the additive property of the trace, we have the following
\begin{equation}
\label{eq:maximum-likelihood-7}
\begin{split}
\big\lbrace 
\widehat{\pmb{\Theta}}^{\tt{tar}},
\widehat{\pmb{\Theta}}^{\tt{u}} 
\big\rbrace
&=
	\argmax_{\lbrace \pmb{\Theta}^{\tt{tar}} , \pmb{\Theta}^{\tt{u}} \rbrace }
	\Big\lbrace
	\sum\nolimits_{k=0}^K 
	\trace 
	\Big(
	\pmb{P}_{\pmb{A}(\pmb{\Theta}^{\tt{tar}}_k)}
	\widehat{\pmb{R}}_k
	\Big)
	\\ & +
	\sum\nolimits_{k=1}^K 
	\frac{\pmb{a}^H(\theta_k^{\tt{u}})
	\pmb{P}_{\pmb{A}(\pmb{\Theta}^{\tt{tar}}_k)}^{\perp}
	\widehat{\pmb{R}}_k
	\pmb{P}_{\pmb{A}(\pmb{\Theta}^{\tt{tar}}_k)}^{\perp} 
	\pmb{a}(\theta_k^{\tt{u}})}{\pmb{a}^H(\theta_k^{\tt{u}})
	\pmb{P}_{\pmb{A}(\pmb{\Theta}^{\tt{tar}}_k)}^{\perp} 
	\pmb{a}(\theta_k^{\tt{u}})
	} 
	\Big\rbrace,
\end{split}
\end{equation}	
The \ac{FML} maximization criterion in \eqref{eq:maximum-likelihood-7} has a nice interpretation. Even though the manifold changes across bands, due to the presence of users at unknown locations, the first term of the \ac{FML} criterion consists of summing the entire data across all the bands. 
Said differently, this term appears due to target-only exploration.The first term also resembles the classical deterministic \ac{ML} criterion of the classical \ac{AoA} estimation problem. On the other hand, the second term appearing in \eqref{eq:maximum-likelihood-7} constitutes the contributions of the $K$ users. 
 In fact, the $k^{th}$ term within the sum implies that estimating $\theta_k^{\tt{u}}$ should be preceded by nulling $\pmb{A}(\pmb{\Theta}^{\tt{tar}}_k)$ via orthogonal projector $\pmb{P}_{\pmb{A}(\pmb{\Theta}^{\tt{tar}}_k)}^{\perp}$. In other terms, a Bartlett-type search is performed to compute $\theta_k^{\tt{u}}$ via $\argmax_{\theta}\frac{\bar{\pmb{a}}^H(\theta) \widehat{\pmb{R}}_k \bar{\pmb{a}}(\theta) }{\Vert \bar{\pmb{a}}(\theta) \Vert^2}$, where $\bar{\pmb{a}}(\theta) = \pmb{P}_{\pmb{A}(\pmb{\Theta}^{\tt{tar}}_k)}^{\perp}\pmb{a}(\theta)$.
  Even more, there are no cross-terms appearing between any two distinct users, which further simplifies the optimization procedure as presented in the next section.

\vspace{-0.1cm}
\section{Fused Maximum Likelihood Estimation}
\label{sec:estimation}
We develop a method based on alternating projections to compute a \ac{FML} estimate of parameters $\pmb{\Theta}^{\tt{tar}}$ and $\pmb{\Theta}^{\tt{u}}$. Indeed, the naive \ac{ML} estimator would perform a brute force search, resulting in a joint $(q+K)$-dimensional search, which is computationally exhaustive. As we shall show, we can transform this multi-dimensional search, by carefully alternating between the spaces spanned by the common manifold and that of users, to a sequence of $1$D searches that alternate until convergence.

\subsection{Initialization}
In the initialization phase of the proposed method, we target optimization of the common manifold term, i.e. the first term in \eqref{eq:maximum-likelihood-7}. Therefore, this phase seeks the following
\begin{equation}
	\label{eq:first-term-maximization}
	\widehat{\pmb{\Theta}}^{\tt{tar}}_1\ldots \widehat{\pmb{\Theta}}^{\tt{tar}}_K
	=
	\argmax_{{\pmb{\Theta}}^{\tt{tar}}_1 \ldots {\pmb{\Theta}}^{\tt{tar}}_K}
	\sum\nolimits_{k} 
	\trace 
	\Big(
	\pmb{P}_{\pmb{A}(\pmb{\Theta}^{\tt{tar}}_k)}
	\widehat{\pmb{R}}_k
	\Big).
\end{equation}
Since this problem corresponds to the classical \ac{AoA} problem, we can adopt the classical alternating projection algorithm described in \cite{7543} to obtain an initial estimate, i.e. $\widehat{\pmb{\Theta}}^{\tt{tar}}_{k,(0)}$. However, even though our main interest lies in the target \ac{AoA}s, this step alone does not suffice, as the resulting estimates $\widehat{\pmb{\Theta}}^{\tt{tar}}_{k,(0)}$ are biased to the presence of the users as per \eqref{eq:maximum-likelihood-7}. 
Notice that since $\lbrace \widehat{\pmb{R}}_k \rbrace_{k=0}^K$ contains all the \ac{AoA} information, namely $\pmb{\Theta}^{\tt{tar}}$, this phase can provide a satisfactory initial estimate of the target \ac{AoA}s.

\subsection{Phase 1: Updating User \ac{AoA}s}
At iteration $p$, one has access to $\widehat{\pmb{\Theta}}^{\tt{tar}}_{k,(p)}$, which is the $p^{th}$ estimate of the target \acp{AoA} within the $k^{th}$ band. This step involves $K$ independent $1-$dimensional searches to obtain $\widehat{\pmb{\Theta}}^{\tt{u}}_{(p)}$, namely, the $k^{th}$ element of $\widehat{\pmb{\Theta}}^{\tt{u}}_{(p)}$ is updated according to the following criterion
\begin{equation}
\label{eq:user-aoa-update-eqn}
	[\widehat{\theta}_k^{\tt{u}}]_{(p)}
	=
	\argmax_{\theta}
	\frac{\pmb{a}^H(\theta)
	\pmb{P}_{
	\pmb{A}(
	\widehat{\pmb{\Theta}}^{\tt{tar}}_{k,(p)}
	)
	}^{\perp}
	\widehat{\pmb{R}}_k
	\pmb{P}_{\pmb{A}(
	\widehat{\pmb{\Theta}}^{\tt{tar}}_{k,(p)}
	)}^{\perp} 
	\pmb{a}(\theta)}{\pmb{a}^H(\theta)
	\pmb{P}_{\pmb{A}(
	\widehat{\pmb{\Theta}}^{\tt{tar}}_{k,(p)}
	)}^{\perp} 
	\pmb{a}(\theta)
	}.
\end{equation}
Since the maximization processes are independent, i.e. the maximization \ac{w.r.t} $\widehat{\theta}_i^{\tt{u}}$ does not depend on $\widehat{\theta}_j^{\tt{u}}$, the $K$ searches can be implemented in parallel.

\subsection{Phase 2: Updating Target \ac{AoA}s}
At iteration number $p$ and after the completion of Phase $1$, we shall use $\widehat{\pmb{\Theta}}^{\tt{tar}}_{k,(p)}$, as well as $\widehat{\pmb{\Theta}}^{\tt{u}}_{(p)}$, to update target \ac{AoA}s and generate their updated estimates. In an attempt to estimate the \ac{AoA} of the $m^{th}$ target, we rewrite \eqref{eq:maximum-likelihood-7} as follows 
\begin{equation}
\label{eq:phase-two-update-target-aoa}
	[\widehat{\theta}_m^{\tt{tar}}]_{(p+1)}
	=
	\argmax_{\theta}
	\sum_k
	\frac{\pmb{a}^H(\theta) \pmb{P}_{\widehat{\pmb{A}}_{k,(p)/m}}^\perp \widehat{\pmb{R}}_k  \pmb{P}_{\widehat{\pmb{A}}_{k,(p)/m}}^\perp  \pmb{a}(\theta)}{\pmb{a}^H(\theta)  \pmb{P}_{\widehat{\pmb{A}}_{k,(p)/m}}^\perp \pmb{a}(\theta)},
\end{equation}
where terms that do not depend on $\theta_m^{\tt{tar}}$ have been omitted. The projector matrix appearing in \eqref{eq:phase-two-update-target-aoa} is obtained by omitting the column corresponding to $\pmb{a}(\theta_m^{\tt{tar}})$ from $\pmb{A}_k$ through
\begin{equation}
\label{eq:Akm-definition}
\widehat{\pmb{A}}_{k,(p)/m}
=
	\begin{cases}
		\pmb{A}(
	\widehat{\pmb{\Theta}}^{\tt{tar}}_{0,(p)/m}
		) &, k = 0 \\
		\begin{bmatrix}
		\pmb{a}([\theta_k^{\tt{u}}]_{(p)})		&
		\pmb{A}(
	\widehat{\pmb{\Theta}}^{\tt{tar}}_{k,(p)/m}
		)
	\end{bmatrix}&, 1 \leq k \leq K,
	\end{cases}
\end{equation}
where $\widehat{\pmb{\Theta}}^{\tt{tar}}_{k,(p)/m}$ is the $p^{th}$ estimate obtained by eliminating the entry corresponding to ${\theta}_m^{\tt{tar}}$ from $\widehat{\pmb{\Theta}}^{\tt{tar}}_{k,(p)}$, and the summation runs on terms where $m \in \Phi_k$.
After iterating through all target \acp{AoA}, we go back to phase $1$, and so on. To this end, we summarize the proposed method in \textbf{Algorithm \ref{alg:alg1}}. It is worth noting that a modified version of this method can be used when one has prior knowledge of $\widehat{\pmb{\Theta}}^{\tt{u}}$ (or at least an estimate of it). This is simply achieved via bypassing phase one of \textbf{Algorithm \ref{alg:alg1}}, and directly utilizing the given $\widehat{\pmb{\Theta}}^{\tt{u}}$ in phase two, i.e. using $\widehat{\pmb{\Theta}}^{\tt{u}}$ in \eqref{eq:Akm-definition} in order to estimate the target \ac{AoA}s. This modification is of interest when users are assumed to be static, for example, an office environment or even static anchors providing regular communication signals towards \ac{BS}.
\subsection{Computational Complexity}
\label{sec:complexity-1}
Notice that the heaviest computational complexity arises from phase $2$ and in particular when updating $\widehat{\pmb{A}}_{k,m}^{(p)}$ via \eqref{eq:Akm-definition}.
Observe that $\pmb{P}_{\widehat{\pmb{A}}_{k,(p)/m}}^\perp$ is independent of $\theta$ but depends on $k$ and $m$. It costs $\mathcal{O}(q^2 N + q^3 + q N^2)$. Summing over all $K$, computing $\pmb{P}_{\widehat{\pmb{A}}_{k,(p)/m}}^\perp$ over all $k$ costs 
	$\mathcal{O}(q^2 K N + q^3 K + q K N^2)$.
	 Also, calculating the terms $\pmb{P}_{\widehat{\pmb{A}}_{k,(p)/m}}^\perp \widehat{\pmb{R}}_k \pmb{P}_{\widehat{\pmb{A}}_{k,(p)/m}}^\perp$ costs $\mathcal{O}(N^3)$. Summing over all $K$ will cost $\mathcal{O}(KN^3)$. Now, each $\theta$ evaluation in the maximization process consists of computing \eqref{eq:phase-two-update-target-aoa}, which costs $\mathcal{O}(KN^2)$. So, the whole search evaluates $\mathcal{O}(N_\theta K N^2)$. Summing over all $q$, we get $\mathcal{O}(qN_\theta K N^2 + q^3 K N + q^4 K + q^2 K N^2 + qKN^3)$, where $N_\theta$ is the grid size.
	 Assuming phase $1$ and phase $2$, as well as the initializer, iterate a total amount of $P_{\tt{iter}}$ times, and noting that phase $2$ dominates phase $1$, the total complexity of \textbf{Algorithm \ref{alg:alg1}} is $
	T_1 = \mathcal{O}\big(P_{\tt{iter}}(qN_\theta K N^2 + q^3 K N + q^4 K + q^2 K N^2 + qKN^3 ) \big)$,   
where we have ignored terms of lower orders.

\begin{algorithm}[H]
\caption{Efficient \ac{FML} estimation}\label{alg:alg1}
\begin{algorithmic}
\STATE \textsc{input}: $\widehat{\pmb{R}}_0 \ldots \widehat{\pmb{R}}_K$
\STATE {\textsc{initialize}:} 
\STATE Run alternating projection method \cite{7543} to maximize first term of \ac{FML} in \eqref{eq:maximum-likelihood-7}, as given in \eqref{eq:first-term-maximization} and set $p \gets 0 $.
\STATE \textsc{do}
\STATE {\tt{\# PHASE 1: Updating User \ac{AoA}s}}
\STATE  \hspace{0.5cm}\textsc{parfor} $ k = 1 \ldots K$
\STATE \hspace{0.5cm} \hspace{0.5cm}Update $[\widehat{\theta}_k^{\tt{u}}]_{(p)}$ via \eqref{eq:user-aoa-update-eqn}.
\STATE {\tt{\# PHASE 2: Updating Target \ac{AoA}s}}
\STATE  \hspace{0.5cm}\textsc{for} $ m = 1 \ldots q$
\STATE \hspace{0.5cm} \hspace{0.5cm}Form $\widehat{\pmb{A}}_{k,m}^{(p)}$ for all $k$ as described in \eqref{eq:Akm-definition}.
\STATE \hspace{0.5cm} \hspace{0.5cm}Update $[\widehat{\theta}_m^{\tt{tar}}]_{(p+1)}$ via \eqref{eq:phase-two-update-target-aoa}.
\STATE $p \gets p + 1$
\STATE \textsc{until} $\Vert \widehat{\pmb{\Theta}}^{\tt{tar}}_{(p+1)} - \widehat{\pmb{\Theta}}^{\tt{tar}}_{(p)} \Vert + \Vert \widehat{\pmb{\Theta}}^{\tt{u}}_{(p+1)} - \widehat{\pmb{\Theta}}^{\tt{u}}_{(p)} \Vert < \epsilon $
\STATE \textbf{return}  $\widehat{\pmb{\Theta}}^{\tt{tar}},\widehat{\pmb{\Theta}}^{\tt{u}}$
\end{algorithmic}
\end{algorithm}

\vspace{-0.3cm}
\section{Fused Subspace Estimation}
\label{sec:estimation-subspace}
In this section, we introduce a novel method tailored for the \ac{HRF} problem at hand. It is known that subspace-based methods are sub-optimal and require less computational complexity than \ac{ML} based ones. First, we revise the principle of \ac{MUSIC} estimation, then we take a step further to formulate an optimization framework inspired by the \ac{MUSIC} principle, via subspace orthogonality tailored for the \ac{HRF} problem.
\subsection{Classical MUSIC estimation}
We start by writing down the expression of the true covariance matrix corresponding to the model given in \eqref{eq:matrix-form-UL} and \eqref{eq:matrix-form-DL}. Indeed, we have 
\begin{equation}
\label{eq:true-covariance-expression}
	\pmb{R}_k
	\triangleq 
	\mathbb{E}
	\big( \pmb{Y}_k \pmb{Y}_k^H \big)
	=
	\pmb{A}_k
	\pmb{R}_{xx}^{(k)}
	\pmb{A}_k^H
	+
	\pmb{R}_{ww,k},
\end{equation}
where $\pmb{A}_k$ is defined in \eqref{eq:Ak-definition}, the covariance matrix $\pmb{R}_{xx}^{(k)}$ is given as $\pmb{R}_{xx}^{(k)} = \mathbb{E}
	\big( \pmb{X}_k \pmb{X}_k^H \big)$, and the covariance matrix $\pmb{R}_{ww,k}$ is the covariance matrix of the \ac{AWGN} which is assumed to be white over different frequency bands, i.e $\pmb{R}_{ww,k} = \sigma^2 \pmb{I}_N$. To this end, the \ac{MUSIC} principle relies on the observation that $\Span(\pmb{A}_k) = \Span(\pmb{R}_k)$, for noiseless scenarios. In fact, when noise is present, we can parametrize the \ac{EVD} of $\pmb{R}_k$ as follows 
\begin{equation}
	\pmb{R}_k
	=
	\pmb{U}_{S,k}
	\pmb{\Lambda}_{S,k}
	\pmb{U}_{S,k}^H
	+
	\pmb{U}_{N,k}
	\pmb{\Lambda}_{N,k}
	\pmb{U}_{N,k}^H.
\end{equation}
The matrices $\pmb{U}_{S,k} $ and $\pmb{U}_{N,k}$ are known to be the \textit{signal} and \textit{noise} subspaces on the $k^{th}$ frequency band, respectively. The dimensions of $\pmb{U}_{S,k} $ are the same as that of $\pmb{A}_k$, and consequently, $\pmb{U}_{N,k}$ would represent an orthogonal complement of $\pmb{U}_{S,k} $. As a result, the eigenvalues are also decomposed into signal and noise eigenvalues contained in the diagonal block matrices $\pmb{\Lambda}_{S,k}$ and $\pmb{\Lambda}_{N,k}$, respectively. Indeed, given that $\pmb{R}_{xx}^{(k)}$ is full-rank, the eigenvalues contained in the $i^{th}$ diagonal entry of $\pmb{\Lambda}_{S,k}$ is $\pmb{\Lambda}_{S,k}(i,i) = \lambda_{k,i} + \sigma^2$, where $\lambda_{k,i}$ is the $i^{th}$ largest eigenvalue of $\pmb{A}_k\pmb{R}_{xx}^{(k)}\pmb{A}_k^H$. Moreover, it is easily verified that $\pmb{\Lambda}_{N,k}(i,i) = \sigma^2$. This "jump" in eigenvalues aid in counting the number of signals within $\pmb{A}_k$. Next, since $\Span(\pmb{U}_{S,k}) = \Span(\pmb{A}_k)$, we have that $\Span(\pmb{U}_{N,k}) \perp \Span(\pmb{A}_k)$.
Therefore, we can guarantee perfect orthogonality between the steering vectors looking towards the true \ac{AoA}s and the noise subspace, i.e. $\pmb{A}_k^H\pmb{U}_{N,k}  = \pmb{0}$, or equivalently $\Vert \pmb{a}^H(\theta)\pmb{U}_{N,k} \Vert^2 = 0$, for all $\theta \in [\theta_k^{\tt{u}},\pmb{\Theta}_k^{\tt{tar}}]$ if $1\leq k \leq K$ and for all $\theta \in [\pmb{\Theta}_0^{\tt{tar}}]$ if $k = 0$.
In practical scenarios, the expectation in \eqref{eq:true-covariance-expression} is replaced with a sample average, as dictated by \eqref{eq:sample-covariance-matrix}. Therefore, the \ac{EVD} operation is directly applied to the sample covariance matrix as follows
\begin{equation}
\label{eq:EVD-on-sample-covariance-matrix}
	\widehat{\pmb{R}}_k
	=
	\widehat{\pmb{U}}_{S,k}
	\widehat{\pmb{\Lambda}}_{S,k}
	\widehat{\pmb{U}}_{S,k}^H
	+
	\widehat{\pmb{U}}_{N,k}
	\widehat{\pmb{\Lambda}}_{N,k}
	\widehat{\pmb{U}}_{N,k}^H.
\end{equation}
Finally, the \ac{MUSIC} criterion estimates the \ac{AoA}s through peak finding of the following cost function
\begin{equation}
\label{eq:MUSIC}
	f^{-1}(\theta) 
	=
	\Vert \pmb{a}^H (\theta)\widehat{\pmb{U}}_{N,k}  \Vert_F^2,
\end{equation}
which turns out to be the \ac{MUSIC} estimator, and the peaks of $f(\theta)$ correspond to the \ac{MUSIC} estimates of $[\theta_k^{\tt{u}},\pmb{\Theta}_k^{\tt{tar}}]$.

\subsection{Proposed Fused \ac{MUSIC} Alternating Projection }
In fact, from an optimization perspective \cite{9052941}, the \ac{MUSIC} criterion can be cast as follows
\begin{equation}
 \label{eq:problem1}
\begin{aligned}
(\mathcal{P}_{\tt{MU}}):
\begin{cases}
\min\limits_{\pmb{\omega}_k(\theta)}&   \Vert \pmb{\omega}_k(\theta) - \pmb{a}(\theta) \Vert^2 \\
\textrm{s.t.}
 &  \pmb{\omega}_k^H(\theta)\widehat{\pmb{U}}_{N,k} = \pmb{0}.
\end{cases}
\end{aligned}
\end{equation}
The solution of \eqref{eq:problem1} is easily verified to be $\pmb{\omega}_k^{\tt{MU}}(\theta) = \widehat{\pmb{U}}_{S,k}\widehat{\pmb{U}}_{S,k}^H\pmb{a}(\theta)$, which when plugged back in the cost function of problem $(\mathcal{P}_{\tt{MU}})$ gives us the \ac{MUSIC} criterion in \eqref{eq:MUSIC}. In this section, we introduce a fused \ac{MUSIC} approach based on alternating projections. The method is iterative by nature and alternates between user and target \ac{AoA}s in a careful manner. In the first phase of the proposed fused method, we leverage the manifold $\pmb{A}(\pmb{\Theta}_0^{\tt{tar}})$ appearing solely in the \ac{DL}. Via this manifold, we first aim at obtaining a set of basis, where each spans the subspaces defined by $\pmb{a}(\theta^{\tt{u}}_1) \ldots \pmb{a}(\theta^{\tt{u}}_K)$. Defining $\bar{\pmb{R}}_k \triangleq \pmb{P}_0^{\perp}
	\pmb{R}_k$ for all $k$, where $\pmb{P}_0^{\perp}$ projects onto the nullspace of the manifold $\pmb{A}(\pmb{\Theta}_0^{\tt{tar}})$, i.e. $\pmb{P}_0^{\perp}\pmb{A}(\pmb{\Theta}_0^{\tt{tar}}) = \pmb{0}$. This orthogonal projector will aid in nulling the effect of the targets explored by the \ac{DFRC} \ac{BS}, while focusing on the user ones. Assuming a diagonal matrix $\pmb{R}_{xx}^{(k)}$, we have 
\begin{equation}
\label{eq:the-first-svd}
	{\bar{\pmb{R}}}_k
	=
	r_{1,1}^{(k)}
	\pmb{P}_0^{\perp}
	\pmb{a}(\theta^{\tt{u}}_k)
	\pmb{a}^H(\theta^{\tt{u}}_k)
	+
	\sigma^2 
	\pmb{P}_0^{\perp}
	=
	\pmb{U}
	\pmb{\Sigma}
	[
		\pmb{v}_{S,k}
		\ \
		\widetilde{\pmb{V}}_{S,k}
	],
\end{equation}
where $r_{1,1}^{(k)}$ is the first element on the diagonal part of $\pmb{R}_{xx}^{(k)}$. Now, applying \ac{SVD} on $\pmb{P}_0^{\perp}
	\pmb{R}_k$, as shown in the second equality, we can extract the singular vector spanning the only remaining vector $\pmb{a}(\theta^{\tt{u}}_k)$ by extracting the left singular vector corresponding to the strongest singular value of $\pmb{\Sigma}$. Note that an estimate of $\pmb{P}_0^{\perp}$ can be obtained via 
\begin{equation}
\label{eq:initial-null-space}
	\widehat{\pmb{P}}_0^{\perp}	
	=
	\pmb{I}_N
	-
	\widehat{\pmb{U}}_{N,0}
	\widehat{\pmb{U}}_{N,0}^H,
\end{equation}	
and, hence, an estimate of $\bar{\pmb{R}}_k$ is obtained as
\begin{equation}
\label{eq:estimated-projected-covariance-phase-one}
	\widehat{\bar{\pmb{R}}}_k
	=
	\widehat{\pmb{P}}_0^{\perp}
	\widehat{\pmb{R}}_k.
\end{equation}
Note that the term $\widehat{\pmb{U}}_{N,0} \in \mathbb{C}^{N \times (N - \vert \Phi_0 \vert)}$ in \eqref{eq:initial-null-space} is obtained through the \ac{EVD} operation, i.e. \eqref{eq:EVD-on-sample-covariance-matrix}, of the sample covariance matrix $\widehat{\pmb{R}}_0$, i.e. \eqref{eq:sample-covariance-matrix} for $k = 0$.
Then, all sample covariance matrices computed via \eqref{eq:sample-covariance-matrix} are pre-multiplied by $\widehat{\pmb{P}}_0^{\perp}$ to obtain $\widehat{\bar{\pmb{R}}}_k$ using \eqref{eq:estimated-projected-covariance-phase-one}.Now that we have $\pmb{v}_{S,k}$, we can define an optimization problem tailored to extract the \ac{AoA} of users. To this end, we propose
\begin{equation}
 \label{eq:problem2}
\begin{aligned}
(\mathcal{P}_{k}):
\begin{cases}
\min\limits_{\omega_k(\theta)}&   \Vert \pmb{v}_{S,k}\omega_k(\theta) - \pmb{a}(\theta) \Vert^2 \\
\textrm{s.t.}
 &  \Big(\pmb{v}_{S,k}\omega_k(\theta) - \pmb{a}(\theta) \Big)^H\pmb{P}_0^{\perp} = \pmb{0}.
\end{cases}
\end{aligned}
\end{equation}
The above optimization problem adjusts $\omega_k(\theta)$ so that $\pmb{v}_{S,k}\omega_k(\theta)$ matches the steering vector $\pmb{a}(\theta)$ in all directions $\theta$ only for vectors residing in the orthogonal complement of $\pmb{A}(\pmb{\Theta}_0^{\tt{tar}})$. More specifically, we seek to minimize the residual $\pmb{v}_{S,k}\omega_k(\theta) - \pmb{a}(\theta)$ for those lying in the orthogonal complement, $\pmb{P}_0^{\perp}$. Accordingly, the associated Lagrangian function of the above optimization problem is given as 
\begin{equation}
\label{eq:lagrangian-problem2}
\begin{split}
	\mathcal{L}(\omega_k(\theta),\pmb{\lambda})
	&=
	\Vert \pmb{v}_{S,k}\omega_k(\theta) - \pmb{a}(\theta) \Vert^2 \\
	&-
	\Big(\pmb{v}_{S,k}\omega_k(\theta) - \pmb{a}(\theta) \Big)^H\pmb{P}_0^{\perp}
	\pmb{\lambda}^{\opt}.
\end{split}
\end{equation} 
Deriving \eqref{eq:lagrangian-problem2} \ac{w.r.t} $\omega_k(\theta)$, we get
\begin{equation}
	\frac{\partial \mathcal{L}(\omega_k(\theta),\pmb{\lambda})}{\partial \omega_k(\theta)}
	=
	2 \pmb{v}_{S,k}^H \Big(\pmb{v}_{S,k}\omega_k(\theta) - \pmb{a}(\theta) \Big) 
	-
	\pmb{v}_{S,k}^H  
	\pmb{P}_0^{\perp}
	\pmb{\lambda}.
\end{equation}
Setting the above gradient to zero and using the property $\pmb{v}_{S,k}^H \pmb{v}_{S,k} = \pmb{I}_N$ $\forall k$, one gets a closed form in $\omega_k(\theta)$ as follows
\begin{equation}
\label{eq:omega-optimal-problem2}
	\omega_k^{\opt}(\theta)
	=
	\pmb{v}_{S,k}^H\pmb{a}(\theta)
	+
	\frac{1}{2}
	\pmb{v}_{S,k}^H
	\pmb{P}_0^{\perp}
	\pmb{\lambda}^{\opt}.
\end{equation}
In order to obtain the optimal value of the Lagrangian multiplier $\pmb{\lambda}^{\opt}$, we derive the Lagrangian \ac{w.r.t} $\pmb{\lambda}$ yielding the following first-order conditions
\begin{equation}
	\pmb{P}_0^{\perp}
	\Big(
	\pmb{v}_{S,k}\pmb{v}_{S,k}^H
	\pmb{a}(\theta)
	+
	\frac{1}{2}
	\pmb{v}_{S,k}\pmb{v}_{S,k}^H
	\pmb{P}_0^{\perp}
	\pmb{\lambda}^{\opt}
	-
	\pmb{a}(\theta)
	\Big)
	=
	\pmb{0}.
\end{equation}
Expanding, then re-arranging, we get
\begin{equation}
\label{eq:optimal-lambda-problem2}
	\pmb{\lambda}^{\opt}
	=
	2
	\big(
	\pmb{P}_0^{\perp}
	\pmb{v}_{S,k}\pmb{v}_{S,k}^H
	\pmb{P}_0^{\perp}
	\big)^+
	\pmb{P}_0^{\perp}
	\big(
	\pmb{I}_N
	-
	\pmb{v}_{S,k}\pmb{v}_{S,k}^H
	\big)
	\pmb{a}(\theta).
\end{equation}
Plugging equation \eqref{eq:optimal-lambda-problem2} in \eqref{eq:omega-optimal-problem2}
\begin{equation}
\label{eq:omega2-optimal-problem2}
\begin{split}
	\omega_k^{\opt}(\theta)
	&=
	\pmb{v}_{S,k}^H
	\pmb{a}(\theta)\\ &
	+ 
	\pmb{v}_{S,k}^H
	\pmb{P}_0^{\perp}
	\big(
	\pmb{P}_0^{\perp}
	\pmb{v}_{S,k}\pmb{v}_{S,k}^H
	\pmb{P}_0^{\perp}
	\big)^+
	\pmb{P}_0^{\perp}
	\big(
	\pmb{I}_N
	-
	\pmb{v}_{S,k}\pmb{v}_{S,k}^H
	\big)
	\pmb{a}(\theta).
\end{split}
\end{equation}
Via property $\pmb{Z}^T(\pmb{Z}\pmb{Z}^T)^+ = \pmb{Z}^+$ \cite{petersen2008matrix}, we can rewrite \eqref{eq:omega2-optimal-problem2}, i.e. 
\begin{equation}
\label{eq:omega3-optimal-problem2}
\begin{split}
	\omega_k^{\opt}(\theta)
	=
	\pmb{v}_{S,k}^H
	\pmb{a}(\theta)
	+
	\big(
	\pmb{P}_0^{\perp}
	\pmb{v}_{S,k}
	\big)^+
	\pmb{P}_0^{\perp}
	\big(
	\pmb{I}_N
	-
	\pmb{v}_{S,k}\pmb{v}_{S,k}^H
	\big)
	\pmb{a}(\theta).
\end{split}
\end{equation}
which, after expanding and by definition of the Moore-Penrose pseudo-inverse, gives
\begin{equation}
\label{eq:omega4-optimal-problem2}
\begin{split}
	\omega_k^{\opt}(\theta)
	&=
	\pmb{v}_{S,k}^H
	\pmb{a}(\theta) +
		\big(
	\pmb{P}_0^{\perp}
	\pmb{v}_{S,k}
	\big)^+
	\pmb{P}_0^{\perp}
	\pmb{a}(\theta) \\
	&-
	\big(
	\pmb{P}_0^{\perp}
	\pmb{v}_{S,k}
	\big)^+
	\pmb{P}_0^{\perp}
	\pmb{v}_{S,k}\pmb{v}_{S,k}^H
	\pmb{a}(\theta)= 
	\big(
	\pmb{P}_0^{\perp}
	\pmb{v}_{S,k}
	\big)^+
	\pmb{P}_0^{\perp}
	\pmb{a}(\theta).
\end{split}
\end{equation}
Finally, plugging \eqref{eq:omega4-optimal-problem2} in the cost function of  \eqref{eq:problem2}, we get
\begin{equation}
\label{eq:first-thetak-spectrum}
	\begin{split}
	g_k(\theta)
	&=
	\big\Vert 
	\big[
	\pmb{v}_{S,k}
	\big(
	\pmb{P}_0^{\perp}
	\pmb{v}_{S,k}
	\big)^+
	\pmb{P}_0^{\perp} - \pmb{I}_N \big]
	\pmb{a}(\theta) 
	\big\Vert^2	\\
	&=
	\big\Vert 
	\big[
	\pmb{v}_{S,k}
	\big(
	\pmb{v}_{S,k}^H
	\pmb{P}_0^{\perp}
	\pmb{v}_{S,k}
	\big)^{-1}
	\pmb{v}_{S,k}^H
	\pmb{P}_0^{\perp} - \pmb{I}_N \big]
	\pmb{a}(\theta) 
	\big\Vert^2. 
	\end{split}
\end{equation}
Now, we can get an estimate of $\widehat{\theta}_k^{\tt{u}}$, which is obtained through computing the maximum of $g_k^{-1}(\theta)$.
The second phase entails in forming $K$ orthogonal projectors $\widehat{\pmb{Q}}_1^{\perp} \ldots \widehat{\pmb{Q}}_K^{\perp}$ as follows
\begin{equation}
	\label{eq:Qk-projector-matrix}
	\widehat{\pmb{Q}}_k^{\perp}
	=
	\pmb{I}_N
	-
	\Vert \pmb{a}(\widehat{\theta}^{\tt{u}}_k) \Vert^{-2}
	\pmb{a}(\widehat{\theta}^{\tt{u}}_k)
	\pmb{a}^H(\widehat{\theta}^{\tt{u}}_k), \quad \forall k = 1 \ldots K
\end{equation}
Next, we define $K$ covariance matrices projected onto $\widehat{\pmb{Q}}_1^{\perp} \ldots \widehat{\pmb{Q}}_K^{\perp}$, respectively, through $\pmb{S}_k
	=
	\pmb{Q}_k^{\perp} 
	\pmb{R}_k$ for all $k \geq 1$. An estimate of $\pmb{S}_k$ is obtained as
\begin{equation}
\label{eq:computing-SK-proj}
	\widehat{\pmb{S}}_k
	=
	\widehat{\pmb{Q}}_k^{\perp} 
	\widehat{\pmb{R}}_k.
\end{equation}  
Using the same process as in the first phase, we can extract an estimate of the subspace spanned by $\pmb{A}(\pmb{\Theta}^{\tt{tar}}_k)$ via the strongest $\vert \Phi_k \vert $ left singular vectors of $\widehat{\pmb{S}}_k$ as follows 
\begin{equation}
\label{eq:obtaining-VSK-thru-Sk}
	\widehat{\pmb{S}}_k
	=
	\pmb{U}_k
	\pmb{\Sigma}_k
	\begin{bmatrix}
		\widehat{\pmb{V}}_{S,k}
		&
		\widetilde{\bar{\pmb{V}}}_{S,k}
	\end{bmatrix},
\end{equation}
where $\widehat{\pmb{V}}_{S,k} \in \mathbb{C}^{N \times \vert \Phi_k \vert}$ and $\Span(\pmb{V}_{S,k}) = \Span(\pmb{A}(\pmb{\Theta}_k^{\tt{tar}}))$ for all $k = 1 \ldots K$. Prior to introducing the optimization problem that aims at estimating $\pmb{\Theta}^{\tt{tar}}$, it is worth noting that all the extracted singular vectors are mere estimates of the span of the common manifold, and so, it is crucial to fuse all the estimated subspaces we have obtained into one optimization problem. For this, we propose problem $(\mathcal{F})$ given as follows
\begin{equation}
 \label{eq:problem3}
\begin{aligned}
(\mathcal{F}):
\begin{cases}
\min\limits_{ \pmb{\omega}_k(\theta) }& 
 \Vert\pmb{\omega}_0(\theta) - \pmb{a}(\theta) \Vert^2 
 + 
\sum\limits_{k=1}^K
 \Vert \widehat{\pmb{V}}_{S,k}\pmb{\omega}_k(\theta) - \pmb{a}(\theta) \Vert^2 \\
\textrm{s.t.}
& \pmb{\omega}_0^H(\theta)\widehat{\pmb{U}}_{N,0} = \pmb{0} \\
 &  \Big(\widehat{\pmb{V}}_{S,k}\pmb{\omega}_k(\theta) - \pmb{a}(\theta) \Big)^H \widehat{\pmb{Q}}_k^{\perp} = \pmb{0}, \ \ \ \ \ \forall k \geq 1.
\end{cases}
\end{aligned}
\end{equation}
We highlight two main aspects of $(\mathcal{F})$ defined in \eqref{eq:problem3}. The first one consists of a simple \ac{MUSIC} criterion in which we aim at fitting a direction vector $\pmb{\omega}_0(\theta)$ orthogonal to $\widehat{\pmb{U}}_{N,0}$. 
The second one entails a joint subspace fit of all $\widehat{\pmb{V}}_{S,k}$ through finding variable weight coefficients $\pmb{\omega}_k(\theta)$ per direction $\theta$ that fits the steering vector in the same direction $\theta$. However, to limit the search space to directions of interest, i.e. those not including the user \acp{AoA}, we constrain each residual to be orthogonal to $\widehat{\pmb{Q}}_k^{\perp}$. The Lagrangian associated with $(\mathcal{F})$ is
\begin{equation*}
\begin{split}
	\mathcal{L}
	&=
\Vert\pmb{\omega}_0(\theta) - \pmb{a}(\theta) \Vert^2
 + 
\sum\nolimits_{k=1}^K
 \Vert \widehat{\pmb{V}}_{S,k}\pmb{\omega}_k(\theta) - \pmb{a}(\theta) \Vert^2 \\
 &-
 \pmb{\omega}_0^H(\theta)\widehat{\pmb{U}}_{N,0}\pmb{\lambda}_0
 -
 \sum\nolimits_{k=1}^K
  \Big(\widehat{\pmb{V}}_{S,k}\pmb{\omega}_k(\theta) - \pmb{a}(\theta) \Big)^H \widehat{\pmb{Q}}_k^{\perp}
  	\pmb{\lambda}_k.
\end{split}
\end{equation*}
The gradients \ac{w.r.t} $\pmb{\omega}_0(\theta),\pmb{\omega}_1(\theta) \ldots \pmb{\omega}_K(\theta)$ are 
\begin{align}
	\frac{\partial \mathcal{L}}{\partial \pmb{\omega}_0(\theta)}
	&=
	2
	(\pmb{\omega}_0(\theta) - \pmb{a}(\theta))
	-
 \widehat{\pmb{U}}_{N,0}\pmb{\lambda}_0, \\
 \frac{\partial \mathcal{L}}{\partial \pmb{\omega}_k(\theta)}
	&=
	2
	\widehat{\pmb{V}}_{S,k}^H
	\big(
	\widehat{\pmb{V}}_{S,k}\pmb{\omega}_k(\theta) - \pmb{a}(\theta) \big)
	-
	\widehat{\pmb{V}}_{S,k}^H
	\widehat{\pmb{Q}}_k^{\perp}
  	\pmb{\lambda}_k.
\end{align}
Setting the gradients to zero, we get the optimal values as 
\begin{align}
	\label{eq:w0opt}
	\pmb{\omega}_0^{\opt}(\theta) 
	&=
	\pmb{a}(\theta)
	+
	\frac{1}{2}
	 \widehat{\pmb{U}}_{N,0}\pmb{\lambda}_0^{\opt}, \\
	 \label{eq:wkopt}
	 \pmb{\omega}_k^{\opt}(\theta)
	 &= 
	 \widehat{\pmb{V}}_{S,k}^H
	 \pmb{a}(\theta)
	 +
	 \frac{1}{2}
	 \widehat{\pmb{V}}_{S,k}^H
	 \widehat{\pmb{Q}}_k^{\perp}
	 \pmb{\lambda}_k^{\opt}.
\end{align}
 Next, we derive the Lagrangian function \ac{w.r.t} $\pmb{\lambda}_0, \pmb{\lambda}_1 \ldots \pmb{\lambda}_K$, and by setting them to zero at the optimal values, we get 
 \begin{align}
 \label{eq:lambda0opt}
 	\frac{\partial \mathcal{L}}{\partial \pmb{\lambda}_0}
 	&=
 	\widehat{\pmb{U}}_{N,0}^H
 	\pmb{\omega}_0^{\opt}(\theta)
 	=
 	\pmb{0}, \\
 	\label{eq:lambdakopt}
 	\frac{\partial \mathcal{L}}{\partial \pmb{\lambda}_k}
 	&=
 	\widehat{\pmb{Q}}_k^{\perp}
 	 \Big(\widehat{\pmb{V}}_{S,k}\pmb{\omega}_k^{\opt}(\theta) - \pmb{a}(\theta) \Big)=
 	 \pmb{0}, \forall k \geq 1.
 \end{align}
 Plugging equations \eqref{eq:w0opt} and \eqref{eq:wkopt} in equations \eqref{eq:lambda0opt} and \eqref{eq:lambdakopt}, respectively, we get
 \begin{align}
 \label{eq:l0-closed-form}
 	\pmb{\lambda}_0^{\opt}
 	&=
 	- 2 \widehat{\pmb{U}}_{N,0}^H \pmb{a}(\theta), \\
 	 \label{eq:lk-closed-form}
 	\pmb{\lambda}_k^{\opt}
 	&=
 	2(\widehat{\pmb{Q}}_k^{\perp}
 	\widehat{\pmb{V}}_{S,k}\widehat{\pmb{V}}_{S,k}^H
 	\widehat{\pmb{Q}}_k^{\perp})^+
 	\widehat{\pmb{Q}}_k^{\perp}
 	( \pmb{I}_N - \widehat{\pmb{V}}_{S,k}\widehat{\pmb{V}}_{S,k}^H )
 	\pmb{a}(\theta). \nonumber
 \end{align}
 First, we replace $\pmb{\lambda}_0^{\opt}$ found in \eqref{eq:l0-closed-form} back in \eqref{eq:w0opt} to get
\begin{equation}
 \label{eq:w0-in-closed-form}
	\pmb{\omega}_0^{\opt}
	(\theta)
	=
	\pmb{a}(\theta)
	-
	\widehat{\pmb{U}}_{N,0} 
	\widehat{\pmb{U}}_{N,0}^H
	\pmb{a}(\theta). 
\end{equation}
Then, we follow similar steps as in \eqref{eq:omega2-optimal-problem2}, \eqref{eq:omega3-optimal-problem2}, \eqref{eq:omega4-optimal-problem2} to get
 \begin{equation}
 \label{eq:wk-in-closed-form}
	\pmb{\omega}_k^{\opt}
	(\theta)
	=
	(\widehat{\pmb{Q}}_k^{\perp}
	\widehat{\pmb{V}}_{S,k})^+ 
	\widehat{\pmb{Q}}_k^{\perp}
	\pmb{a}(\theta)
	,
	\quad
	\forall k = 1 \ldots K.
\end{equation}
Furthermore, we replace the quantities $\pmb{\omega}_0^{\opt}, \pmb{\omega}_1^{\opt} \ldots \pmb{\omega}_K^{\opt}$ in the cost function of problem  \eqref{eq:problem3}. So, estimating $\pmb{\Theta}^{\tt{tar}}$ involves a peak finding criterion of $q$ \ac{AoA}s via the following $1$D search
\begin{equation}
\label{eq:phase-two-Theta}
\widehat{\pmb{\Theta}}^{\tt{tar}}
=
	\argmax\nolimits_{\theta} h^{-1}(\theta),
\end{equation} 
where $h(\theta)$ is given by
 \begin{equation}
 \label{eq:h-theta-expression-for-algo-2}
 h(\theta)
 =
 \Vert
\widehat{\pmb{U}}_{N,0}^H
	\pmb{a}(\theta) 
\Vert^2
 + 
\sum\limits_{k=1}^K
 \Vert \widehat{\pmb{V}}_{S,k}(\widehat{\pmb{Q}}_k^{\perp}
	\widehat{\pmb{V}}_{S,k})^+ 
	\widehat{\pmb{Q}}_k^{\perp}
	\pmb{a}(\theta) - \pmb{a}(\theta) \Vert^2.
 \end{equation}
Notice that the first quantity of $h(\theta)$ is nothing other than the \ac{MUSIC} cost function, while the $k^{th}$ term of the second quantity is the residual after projecting onto the null-space defined through $\widehat{\pmb{Q}}_k^{\perp}$. Now that we have an estimated value of $\widehat{\pmb{\Theta}}^{\tt{tar}}$, we can update the orthogonal projector $\widehat{\pmb{P}}_0^{\perp}$ as $\widehat{\pmb{P}}_0^{\perp}
	=
	\pmb{I}_N
	-
	\pmb{A}(\widehat{\pmb{\Theta}}^{\tt{tar}})
	\big(
	\pmb{A}^H(\widehat{\pmb{\Theta}}^{\tt{tar}})
	\pmb{A}(\widehat{\pmb{\Theta}}^{\tt{tar}})
	\big)^{-1}
	\pmb{A}^H(\widehat{\pmb{\Theta}}^{\tt{tar}})$.
Finally, we go back to the first phase of the method and repeat until convergence. A summary is given in \textbf{Algorithm \ref{alg:alg2}}.

\subsection{Computational Complexity}
\label{sec:complexity-2}
Phase $2$ of \textbf{Algorithm \ref{alg:alg2}} is the most intensive block of the algorithm and its computational complexity is analyzed as follows: For each $\theta$, the computation of $h(\theta)$ in \eqref{eq:h-theta-expression-for-algo-2} involves the computation of $ \Vert
\widehat{\pmb{U}}_{N,0}^H
	\pmb{a}(\theta) 
\Vert^2$ which costs $\mathcal{O}(N(N-q))$, followed by the computation of $\widehat{\pmb{V}}_{S,k}(\widehat{\pmb{Q}}_k^{\perp}
	\widehat{\pmb{V}}_{S,k})^+ 
	\widehat{\pmb{Q}}_k^{\perp}$ which costs $\mathcal{O}(qN^2 + q^2 N + q^3 + N^3)$, and is computed once per $k$ before grid evaluation. 
	Therefore, summing over all $K$, the computations of all $\widehat{\pmb{V}}_{S,k}(\widehat{\pmb{Q}}_k^{\perp}
	\widehat{\pmb{V}}_{S,k})^+ 
	\widehat{\pmb{Q}}_k^{\perp}$ over all $k$ costs $\mathcal{O}(KqN^2 + Kq^2 N + Kq^3 + KN^3)$.
	The evaluation of $\sum_{k=1}^K
 \Vert \widehat{\pmb{V}}_{S,k}(\widehat{\pmb{Q}}_k^{\perp}
	\widehat{\pmb{V}}_{S,k})^+ 
	\widehat{\pmb{Q}}_k^{\perp}
	\pmb{a}(\theta) - \pmb{a}(\theta) \Vert^2$ costs $\mathcal{O}(KN^2)$
	In short, the dominating terms of this step is $\mathcal{O}(N_\theta K N^2 + KqN^2 + Kq^2 N + Kq^3 + KN^3) $.
	Assuming the algorithm runs for $P_{\tt{iter}}$ iterations, the total complexity is $T_2 = \mathcal{O}\big(P_{\tt{iter}}(
	N_\theta K N^2 + KqN^2 + Kq^2 N + Kq^3 + KN^3
	) \big)$. In short, we observe that  $\frac{T_1}{T_2} \simeq \mathcal{O}(q)$.

\begin{algorithm}[H]
\caption{Alternating "Project-then-Fuse" Subspace Estimation}\label{alg:alg2}
\begin{algorithmic}
\STATE \textsc{input}: $\widehat{\pmb{R}}_0 \ldots \widehat{\pmb{R}}_K$
\STATE {\textsc{initialize}:} 
\STATE $0.1)$ Get the noise subspace of \ac{DL} covariance $\widehat{\pmb{U}}_{N,0}$ by \ac{EVD} on $\widehat{\pmb{R}}_0$ as in \eqref{eq:EVD-on-sample-covariance-matrix}.
\STATE $0.2)$ Compute  $\widehat{\pmb{P}}_0^{\perp}	
	\gets
	\pmb{I}_N
	-
	\widehat{\pmb{U}}_{N,0}
	\widehat{\pmb{U}}_{N,0}^H$ 
\STATE {\tt{\# PHASE 1: Project then estimate.}}
\STATE  \textsc{for} $ k = 1 \ldots K$
\STATE \hspace{0.5cm} $1.1)$ Obtain $\widehat{\bar{\pmb{R}}}_k 
$ via \eqref{eq:estimated-projected-covariance-phase-one}.
\STATE \hspace{0.5cm} $1.2)$ Extract $\pmb{v}_{S,k}$ via \ac{SVD}  as shown in \eqref{eq:the-first-svd}.
\STATE \hspace{0.5cm} $1.3)$ Obtain $\widehat{\theta}_k^{\tt{u}}$ by maximizing \eqref{eq:first-thetak-spectrum}.
\STATE \hspace{0.5cm} $1.4)$ Update $\widehat{\pmb{\Theta}}_k^{\tt{u}} \gets \widehat{\theta}_k^{\tt{u}}$.
\STATE \hspace{0.5cm} $1.5)$ Compute $\widehat{\pmb{Q}}_k^{\perp}$, given $\widehat{\theta}_k^{\tt{u}}$ via \eqref{eq:Qk-projector-matrix}.
\STATE \hspace{0.5cm} $1.6)$ Compute $\widehat{\pmb{S}}_k$ following \eqref{eq:computing-SK-proj}.
\STATE \hspace{0.5cm} $1.7)$ Extract $\widehat{\pmb{V}}_{S,k}$ through \ac{SVD} as in \eqref{eq:obtaining-VSK-thru-Sk}.
\STATE {\tt{\# PHASE 2: Fuse then estimate.}}
\STATE $2.1)$ Get $\widehat{\pmb{\Theta}}^{\tt{tar}}$ via peak finding of $h^{-1}(\theta)$ as in \eqref{eq:phase-two-Theta}.
\STATE $2.2)$ Update $\widehat{\pmb{P}}_0^{\perp}$ using the most recent estimates of $\widehat{\pmb{\Theta}}^{\tt{tar}}$ as $\widehat{\pmb{P}}_0^{\perp}
	=
	\pmb{I}_N
	-
	\pmb{A}(\widehat{\pmb{\Theta}}^{\tt{tar}})
	\big(
	\pmb{A}^H(\widehat{\pmb{\Theta}}^{\tt{tar}})
	\pmb{A}(\widehat{\pmb{\Theta}}^{\tt{tar}})
	\big)^{-1}
	\pmb{A}^H(\widehat{\pmb{\Theta}}^{\tt{tar}}).$ 
\STATE $2.3)$ Go back to Phase $1$.
\STATE \textbf{return}  $\widehat{\pmb{\Theta}}^{\tt{tar}},\widehat{\pmb{\Theta}}^{\tt{u}}$
\end{algorithmic}
\end{algorithm}


\vspace{-0.3cm}
\section{Hybrid Radar Fusion with Directional Beams}
\label{sec:directional-beams}
In this section, we focus on directional beamforming in order to optimize beams dedicated for \ac{SnC} tasks.
Indeed, the use of omni-directional beamforming greatly deteriorates the \ac{DL} detection performance.
For this, we denote the set $\pmb{\Omega}_r$ as the radar region, i.e. the region where the \ac{DFRC} \ac{BS} intends to explore targets. Let's assume that $
\pmb{\Omega}_r$ is centered around direction $\theta_r$ with a main-lobe configured through $\Delta_r$, i.e. 
 $\pmb{\Omega}_r = [\theta_r - \Delta_r,\theta_r + \Delta_r]$.
Note that this is of particular interest when the \ac{DFRC} \ac{BS} is tuned to estimate target \ac{AoA}s that fall within a certain sector $\pmb{\Omega}_r$.
On the other hand, we denote the sets $\pmb{\Omega}_c^{(k)}$ as the region beaming towards the $k^{th}$ user for \ac{DL} communications, i.e. one choice can be to select $\pmb{\Omega}_c^{(k)} = [\theta_k^{\tt{u}} - \Delta_k^{\tt{u}},\theta_k^{\tt{u}} + \Delta_k^{\tt{u}}]$. 
For such a choice, $\Delta_k^{\tt{u}}$ represents the edge of the main-lobe towards the $k^{th}$ communication user. 

\begin{figure}[!t]
\centering
\includegraphics[width=3in]{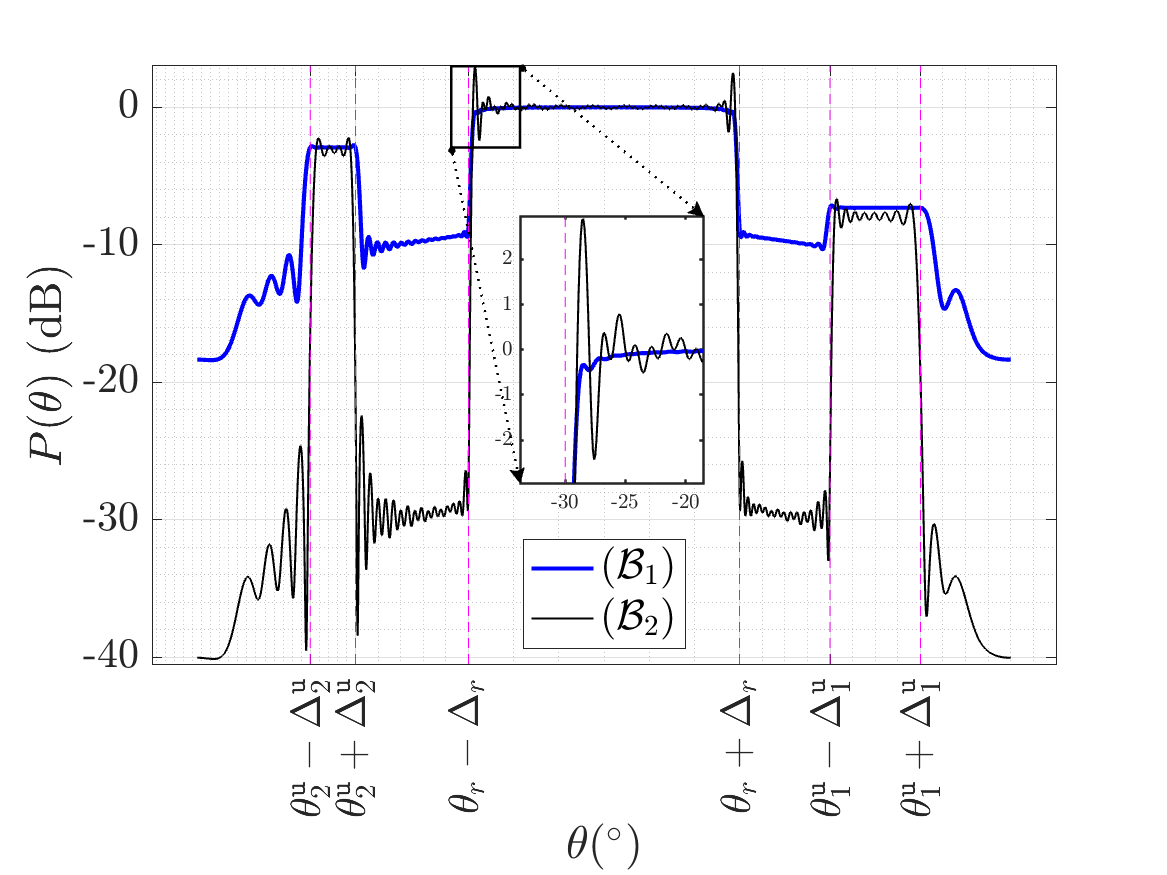}
\caption{The resulting normalized beampatterns of $(\mathcal{B}_1)$ and $(\mathcal{B}_2)$.
The radar operation is centered at $\theta_r = 0^\circ$ with $\Delta_r = 30^\circ$.
The communication users are located at $\theta_1^{\tt{u}} = 60^\circ$ and $\theta_2^{\tt{u}} = -60^\circ$.
The problems were configured to beam-widths of $\Delta_1^{\tt{u}} = 10^\circ$ and $\Delta_2^{\tt{u}} = 5^\circ$.
The power allocated for radar operation is $\varphi_r = 0 \dB$.
The power allocated for users $1$ and $2$ correspond to $\varphi_1 = -3\dB$ and $\varphi_2 = -7.5\dB$, respectively.
}
\label{fig_beamdesign}
\end{figure}

Therefore, a suitable optimization problem tailored for the \ac{HRF} \ac{SnC} directional beam-design problem can be translated as the following $\min-\max$ constrained problem 
\begin{equation}
 \label{eq:problem1}
\begin{aligned}
(\mathcal{B}_1):
\begin{cases}
\min\limits_{\pmb{R}_{\pmb{f}}}\max\limits_{\theta \in \bar{\pmb{\Omega}} } & \pmb{a}_t^H(\theta) \pmb{R}_f \pmb{a}_t(\theta)  \\
\textrm{s.t.}
 &  \pmb{a}_t^H(\theta) \pmb{R}_f \pmb{a}_t(\theta) = \varphi_k, \quad \forall \theta \in \pmb{\Omega}_c^{(k)}, \forall k \\
 &  \pmb{a}_t^H(\theta) \pmb{R}_f \pmb{a}_t(\theta) = \varphi_r, \quad \forall \theta \in \pmb{\Omega}_r \\
 & \trace(\pmb{R}_f) \leq P, \quad \pmb{R}_f  \succeq \pmb{0},
\end{cases}
\end{aligned}
\end{equation}
where $\bar{\pmb{\Omega}} \cup \pmb{\Omega}_r \cup \pmb{\Omega}_c^{(1)} \cup \ldots \cup \pmb{\Omega}_c^{(K)} $ covers the entire angular region. Therefore, the region $\bar{\pmb{\Omega}}$ is a non-\ac{ISAC} region. 
In the above \ac{ISAC} beam-design problem, the covariance matrix $\pmb{R}_f = \mathbb{E}(\pmb{f}\pmb{f}^H)$ is the covariance of the transmit beamformer, which is optimized to minimize the non-\ac{ISAC} beam emissions, in the worst-case sense. This optimization occurs under the following \ac{ISAC} constraints: the $k^{th}$ communication user is allocated a main-lobe power of $\varphi_k$, whereas the radar beam is allocated a main-lobe power of $\varphi_r$. A power budget $P$ is also considered.
Note that this optimization assumes the knowledge of $\theta_k^{\tt{u}}$, which can be obtained by running \textbf{Algorithm \ref{alg:alg1}} or \textbf{Algorithm \ref{alg:alg2}}.
Even more, the beam-width towards the $k^{th}$ user, defined by $\Delta_k^{\tt{u}}$, accounts for location uncertainty around the $k^{th}$ user, which may arise due to location estimation errors.
 Furthermore, these estimates can be used to further optimize $\pmb{f}$ based on problem $(\mathcal{B}_1)$ in \eqref{eq:problem1} to enhance the \ac{DL} detection performance.
Problem $(\mathcal{B}_1)$ is a convex optimization problem and can be solved with off-the-shelf solvers, such as CVX. Even though $(\mathcal{B}_1)$ can provide flat beams for \ac{RnC}, however, it may not be able to minimize out-of-band emissions, i.e. angles within $\bar{\pmb{\Omega}}$, to a satisfactory level. Therefore, another \ac{ISAC} beam-design we can consider is through minimizing an \ac{LS} criterion given a desired spectral mask. More specifically, consider an \ac{ISAC} mask $m(\theta)$ that is defined as follows 
\begin{equation}
\label{eq:m-theta}
	m(\theta)
	=
	\begin{cases}
		\varphi_r, & \quad \text{if } \theta \in \pmb{\Omega}_r \\
		\varphi_k, & \quad \text{if } \theta \in \pmb{\Omega}_c^{(k)}, \forall k = 1 \ldots K \\
		0, & \quad \text{elsewhere}.
	\end{cases}
\end{equation}
Now, given the mask $m(\theta)$ defined in \eqref{eq:m-theta}, we can design beams that are close to $m(\theta)$ in the least-squared sense, i.e.
\begin{equation}
\begin{aligned}
(\mathcal{B}_2):
\begin{cases}
\min\limits_{\pmb{R}_{\pmb{f}}, \alpha}& \sum\nolimits_\theta \vert \pmb{a}_t^H(\theta) \pmb{R}_f \pmb{a}_t(\theta) - \alpha m(\theta) \vert^2  \\
\textrm{s.t.}
 & \trace(\pmb{R}_f) \leq P, \quad \pmb{R}_f  \succeq \pmb{0}.
\end{cases}
\end{aligned}
\end{equation}
Note that $\alpha$ is just a scaling parameter. Problem $(\mathcal{B}_2)$ is also a convex optimization problem and can be solved with off-the-shelf solvers. As an example, we show the resulting normalized beampatterns in Fig. \ref{fig_beamdesign}.
We can observe that problem $(\mathcal{B}_2)$ can minimize out-of-band emissions to as low as $-40\dB$, whereas it is $-18\dB$ for $(\mathcal{B}_1)$. Such out-of-band emission minimization comes at the price of added ripple effect appearing within the radar, as well as the communication beams. For instance, referring to Fig. \ref{fig_beamdesign}, the beam-pattern corresponding to problem $(\mathcal{B}_2)$ ripples around $-2.4\dB$ and $2.9\dB$ at the edge of the radar sector then stabilizes around $0\dB$ as the look-direction approaches the center of the radar region.
In practical scenarios, directional beams are configured to enhance the probability of detection. 
As a result, the \ac{DFRC} \ac{BS} sweeps the environment utilizing beam-scanning techniques for detection.
Denoting $p$ as the beam index, then the corresponding radar region becomes $\pmb{\Omega}_r^{(p)} = [\theta_r^{(p)} - \Delta_r,\theta_r^{(p)} + \Delta_r]$, where $\theta_r^{(p)}$ is the center of the $p^{th}$ radar region. 
Given region $\pmb{\Omega}_r^{(p)}$, both beam design problems $(\mathcal{B}_1)$ and $(\mathcal{B}_2)$ can be solved as discussed.

\vspace{-0.5cm}
\section{Discussion}
\label{sec:discussion}
The reflected signal typically experiences a significant decrease in power due to double path-loss, consequently restricting detection performance unless dealt with pragmatically in practice.
An area of active research involves integrating an \ac{ADC} model, such as $1$-bit or few-bit, that maps the analog signal, i.e. equation \eqref{eq:rx-OFDM-UL}, to digital, i.e. \eqref{eq:sample-rx-OFDM} via an \ac{ADC} function $\mathcal{Q}(.)$; see \cite{8171203} and \cite{mezghani2018blind} for more details. 
Future investigations will delve into the estimation and detection limits with quantized received signals in order to examine the limitations of \ac{HRF} when the reflected powers are significantly smaller than the direct \ac{LoS}.
Regardless of the scenario, the development of advanced algorithms specialized for low-resolution \acp{ADC} is essential to resolve these paths (see \cite{9054740} for a proof-of-concept with $1-$bit \ac{ADC}).
Besides reflection powers versus user powers, future work will also accommodate power control mechanisms tailored for \ac{HRF} as strong users should not overwhelm weaker users, which is of good advantage for low-resolution \ac{ADC} structures \cite{7894211}.
As will be shown in the simulations section, the \ac{HRF} has the potential of improving \ac{MSE} performance on \ac{AoA}, and in particular over mono-static communication-centric \ac{ISAC}, as well as bi-static. This has a desirable impact not only on target sensing, but also on providing better channel estimates, which can be re-used for communication purposes. However, since we provide only a partial view of \ac{UL} channels (i.e. delay, Doppler and gain are treated as nuisance parameters), then additional estimators are needed to compute the required sensing parameters. Hence, an overall improvement in \ac{MSE} quality of sensing parameters has a direct impact on providing a better channel estimate, which has favorable returns on the decoding process for communications.

\section{Simulation Results}
\label{sec:simulations}
\subsection{Parameter Setup}
\label{subsec:parameter-setup}
Assume that both the transmit and receive antenna arrays are equipped with $N$ antennas and follow a \ac{ULA} configuration spaced at half a wavelength, i.e. $\frac{\lambda}{2}$. The carrier frequency is set at $f_c = 24$GHz. For \ac{OFDM} transmissions, the subcarrier spacing is set to $\Delta_f = 240$kHz; hence, the symbol duration is $T = \frac{1}{\Delta_f} = 4.1667 \mu$sec. The cyclic prefix duration is set to $T_{\text{CP}} = \frac{T}{4} = 1.0417 \mu$sec. Therefore, the total symbol duration is equal to $T_o = T + T_{\text{CP}} = 5.2084\mu$sec. The constellation used is a \ac{QPSK}. 
Users are assumed to be located at equidistant angles, namely, the $k^{th}$ user is located at $\theta_k^{\tt{u}} = [-10 - (k-1) \frac{60}{K-1} ]^{\circ}$. All users are considered to be in \ac{LoS} with the \ac{DFRC} \ac{BS}. 
Monte Carlo type experiments are conducted herein, where each trial generates an independent channel realization. The number of Monte Carlo simulations is set to $10^4$ trials.  
The \ac{MSE} is computed as $\MSE =\frac{1}{qE} \sum_{e=1}^E\sum_{p=1}^q \big( \widehat{\theta}_p^{\tt{tar}}(e) - \theta_k \big)^2$, where $ \widehat{\theta}_p^{\tt{tar}}(e)$ is the estimate of the $p^{th}$ target in the $e^{th}$ Monte Carlo experiment, and $E$ is the total number of experiments. All other parameters are mentioned in Section \ref{subsec:simulations}.

The system setting adopted sets $d_k^{\tt{u}} = 100$m for all users. The distances between the $k^{th}$ user and the $i^{th}$ target, i.e. $d_{k,i}$, are chosen from the $(k,i)^{th}$ entry of the following matrix 
\begin{equation}
   \pmb{D}
   =
   \begin{bmatrix}
   64.8489  &  77.7906 &    94.2109\\
   68.6911  &  84.9746 &  100.8081\\
   74.9821 &   92.8385 &  107.1580\\
   82.8447 &  100.8129 &  112.9554 	
   \end{bmatrix}
   m.
\end{equation}
On the other hand, the distance between the $i^{th}$ target and the \ac{DFRC} \ac{BS}, i.e. $d_i^{\tt{tar}}$ is selected from the $i^{th}$ entry of $i^{th}$ element of the following vector $\pmb{d}^{\tt{tar}}
	= \begin{bmatrix} 36 & 32.7951 & 27.0933 \end{bmatrix}m$. Simulations consider either $1$ or $3$ targets. When $1$ target is selected, the index $i = 1$ is used. The same approach is considered for users.
The receive power is selected in a way to match the \ac{SNR} for \ac{MSE} evaluation.

\subsection{Benchmark Schemes}
\label{subsec:benchmark-schemes}
For the sake of performance comparisons, we include the following benchmarks in our performance study:
\paragraph{Naive \ac{MUSIC}} We coin the term \textit{"Naive MUSIC"}, as the method which simply averages all the covariance matrices at hand to form a single covariance matrix $\widehat{\pmb{R}}=\frac{1}{K} \sum_{k=1}^K \widehat{\pmb{R}}_k$. This step is followed by an \ac{EVD} to extract the signal and noise subspaces in order to compute the \ac{MUSIC} criterion. 
\paragraph{\textbf{Algorithm \ref{alg:alg1}} with prior knowledge of user \ac{AoA}s} This benchmark runs \textbf{Algorithm \ref{alg:alg1}} under the condition that the true user \ac{AoAs} are fed to the algorithm. More specifically, the first phase of \textbf{Algorithm \ref{alg:alg1}} is then unnecessary, and is therefore bypassed, because $\widehat{\theta}_k^{\tt{u}}$ are replaced by ${\theta}_k^{\tt{u}}$, for all $k = 1 \ldots K$.
In practice, prior information can be leveraged with the aid of \ac{GNSS} \cite{9013374}.
In \cite{7536855}, the authors used high-precision accuracy of $1$m. Note that centimeter and millimeter level accuracy is also possible \cite{Odijk2016}. An alternative is to run \textbf{Algorithm \ref{alg:alg1}} or \textbf{Algorithm \ref{alg:alg2}} without prior knowledge of user positions in a first stage, then utilize the estimated user positions in the second stage.

\paragraph{\ac{CRB}} The \ac{CRB} is a benchmark giving the lower bound on the variance of any unbiased estimator of deterministic parameters. Following the model introduced in Section \ref{sec:system-model}, the \ac{CRB} of the target \ac{AoAs} \cite{17564} can be shown to be
\begin{equation}
	\label{eq:HRF-CRB}
	\CRB(\pmb{\Theta}^{\tt{tar}}) = \frac{\sigma^2}{2}
	\Big(
    \sum\nolimits_{k=0}^K
	\sum\nolimits_{\ell=0}^{\bar{L}_k}
	\pmb{S}_k \pmb{F}_{k,\ell} \pmb{S}_k^T
	\Big)^{-1},
\end{equation}
where
\vspace{-0.25cm}
\begin{equation}
	\pmb{F}_{k,\ell}
	=
	\Real
	 \lbrace \diag(\pmb{X}_{k}(:,\ell))^* \pmb{H}_k \diag(\pmb{X}_{k}(:,\ell)) \rbrace, 
\end{equation}
\begin{equation}
	\pmb{H}_k
	=
	\pmb{D}_k^H
	(
	\pmb{I}
	-
	\pmb{P}_k
	)
	\pmb{D}_k.
\end{equation}
The expression of $\pmb{P}_k$ is given in \eqref{eq:the-proj-matrix}, $\pmb{D}_k$ is the derivative of the manifold matrix given in \eqref{eq:Ak-definition}, and $\pmb{S}_k$ is a selection matrix indexing columns of the common manifold matrix, $\pmb{A}({\pmb{\Theta}}^{\tt{tar}})$. Also, the \ac{CRB} of the monostatic and bistatic cases (for $K=1$) are also illustrated to show the advantage of the hybrid scheme. The communication-centric mono-static \ac{ISAC} \ac{CRB} accounts only for the \ac{DL}, hence it contains only the $k=0$ term.
The communication-centric bi-static \ac{ISAC} \ac{CRB}, on the other hand, incorporates the \ac{UL} terms, i.e. the $k=1 \ldots K$ terms.

\vspace{-0.5cm}
\subsection{Scenarios}
We distinguish between two scenarios:
\begin{itemize}
	\item Scenario $1$: Here, the product of the number of subcarriers and the number of \ac{OFDM} symbols are set to be equal. In other words, for any $K$ users, the factor $L_k \vert \mathcal{C}_k \vert $ is held constant for all $k$. The aim of studying this scheme is to show the performance of time-frequency resource fairness and its impact on the \ac{MSE} of the estimated \ac{AoA}s via the different methods described in this article. Specifically, if $K$ users are participating in the \ac{UL}, the $k^{th}$ user is allocated $\vert \mathcal{C}_k \vert = \frac{32}{K} $ subcarriers, and $L_k = 16K$ symbols.  
	\item Scenario $2$: In this scenario, we fix the number of \ac{OFDM} symbols to $L_k = 32$ for each users and allocate $\vert \mathcal{C}_k \vert = 32 $ subcarriers for each user. In other words, more users participating in the \ac{UL} requires more bandwidth. 
\end{itemize}
\vspace{-0.5cm}
\subsection{Simulation Results}
\label{subsec:simulations}
\begin{figure*}[!t]
\centering
\subfloat[$N = 5$ antennas]{\includegraphics[width=2.5in]{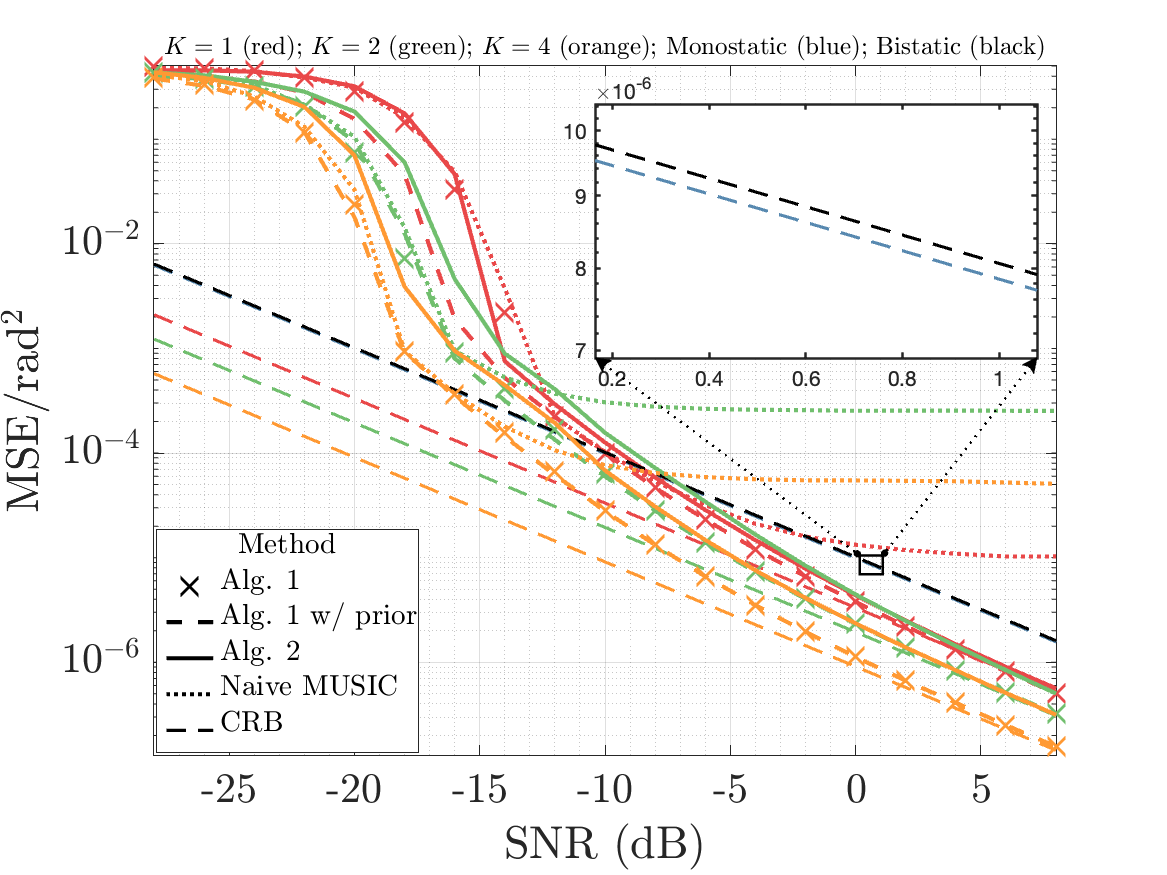}%
\label{fig:scene1_1_q1}}
\hfil
\subfloat[$N = 10$ antennas]{\includegraphics[width=2.5in]{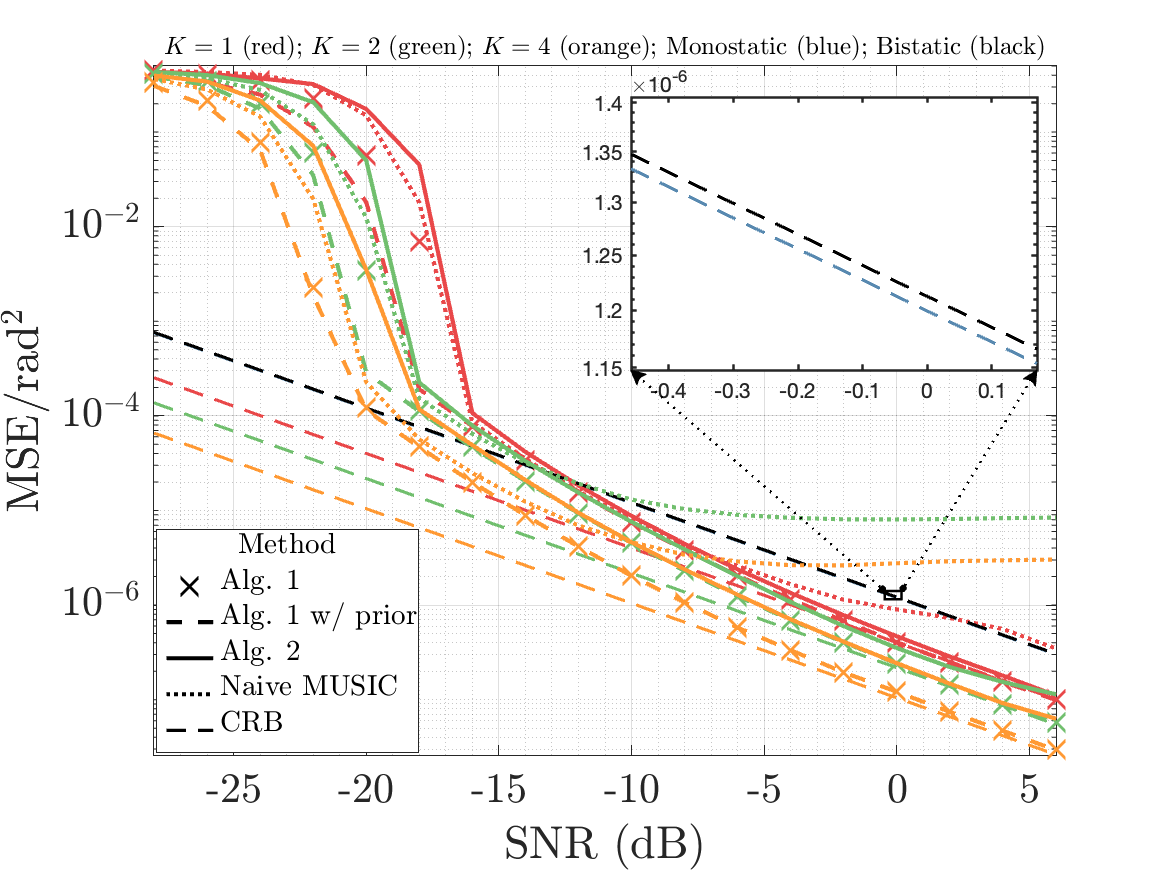}%
\label{fig:scene1_2_q1}}
\caption{Scenario 1: The \ac{MSE} versus \ac{SNR} of the proposed methods and benchmarks for different number of users and $q=1$ target located at $0^\circ$.}
\label{fig:scene1_q1}
\vspace{-0.4cm}
\end{figure*}
\vspace{-0.1cm}
\paragraph{Scenario $1$ with a single target}
In Fig. \ref{fig:scene1_q1}, we plot the \ac{MSE} of the estimated \ac{AoA}s as a function of varying \ac{SNR}, for different number of antennas. 
In Fig. \ref{fig:scene1_1_q1}, we fix $N =5$ antennas and $q=1$ target located at $\theta_1^{\tt{tar}} = 0^\circ$. We see that \textbf{Algorithm \ref{alg:alg1}} with \ac{AoA} prior achieves the best performance for any $K$ as compared to other methods, which is caused by less unknown parameters being estimated than other presented methods. Interestingly, as $K$ increases, \textbf{Algorithm \ref{alg:alg1}} without the user \ac{AoA} prior knowledge coincides with that with prior information. Moreover, both versions of \textbf{Algorithm \ref{alg:alg1}} achieve the \ac{CRB} starting at an \ac{SNR} of $2\dB$. On the other hand, for an \ac{MSE} accuracy of $10^{-4}$, \textbf{Algorithm \ref{alg:alg2}} loses about $4\dB$ in \ac{SNR} as compared to \textbf{Algorithm \ref{alg:alg1}}, and outperforms the Naive \ac{MUSIC} benchmark, even at high \ac{SNR}. When more users are participating in the \ac{UL}, we see that doubling the number of users contributes to an overall \ac{MSE} improvement of about $3.2\dB$. This can be explained through time diversity, i.e. the participation of more users within the same bandwidth, but through the availability of more \ac{OFDM} symbols. 
In Fig. \ref{fig:scene1_2_q1}, we fix $N =10$ antennas, as well as $q=1$ target with $\theta_1^{\tt{tar}} = 0^\circ$. We can notice that both versions of \textbf{Algorithm \ref{alg:alg1}} achieve the \ac{CRB} for all $K$, with an improved performance over that in Fig. \ref{fig:scene1_1_q1} due to an increase in number of antennas. Moreover, as more users participate in the \ac{UL}, the performance of both versions of \textbf{Algorithm \ref{alg:alg1}} approaches its \ac{CRB} faster. This is due to the participation of more users in the \ac{UL} over the same bandwidth resource. For instance, at $\SNR = -20\dB$, \textbf{Algorithm \ref{alg:alg1}} is only $0.0017\rad^2$ far away from its \ac{CRB} for $K=4$, and $0.18\rad^2$ for $K=1$. Moreover, we see that for increasing $K$, \textbf{Algorithm \ref{alg:alg1}} approaches the version of itself with prior user \ac{AoAs}, which is explained through time diversity, i.e. when more \ac{OFDM} symbols are available at the disposal of the \ac{DFRC} \ac{BS}, a better \ac{MSE} performance is expected, in general. In particular, for an accuracy of $10^{-3}$, and for $K = 1$, the gap between both versions of \textbf{Algorithm \ref{alg:alg1}} is $1.2\dB$, compared to $1\dB$ for $K=2$, and $0.7\dB$ for $K =4$. \textbf{Algorithm \ref{alg:alg2}} generally performs better than the Naive \ac{MUSIC} benchmark in the high \ac{SNR} regime. For example, at $K=2$, the performance of \textbf{Algorithm \ref{alg:alg2}} begins to outperform the Naive \ac{MUSIC} benchmark starting from $\SNR = -11\dB$, and $\SNR = -7\dB$ for $K = 8$, which can also be explained through more users transmitting in the \ac{UL}, thus providing more \ac{OFDM} samples towards the \ac{DFRC} \ac{BS}. We notice that for both $N=5$ and $N = 10$ cases, the \ac{CRB} of the mono-static and bi-static are very close. In fact, the \ac{HRF} \ac{CRB} for only $K = 1$ user, improves the \ac{MSE} by nearly $5\dB$.
\begin{figure*}[!t]
\centering
\subfloat[$N = 5$ antennas]{\includegraphics[width=2.5in]{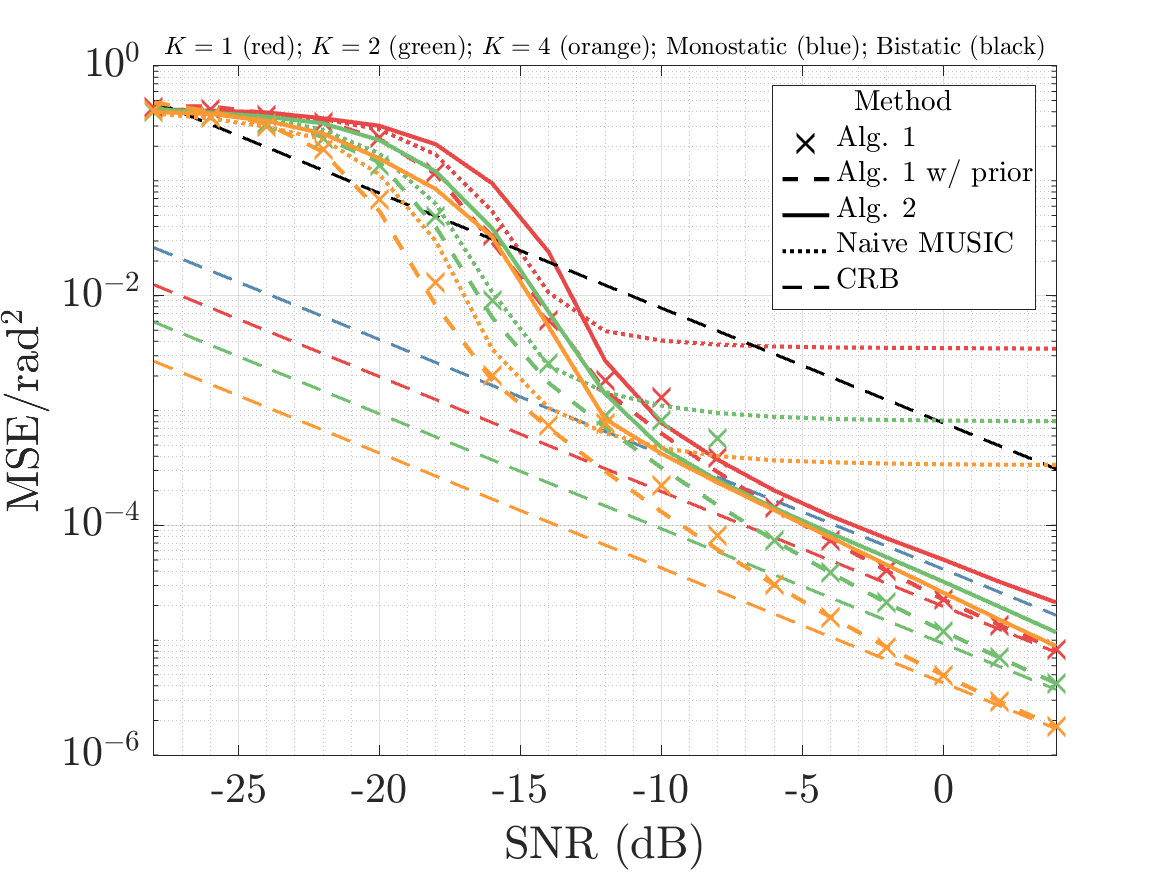}%
\label{fig:scene1_1_q3}}
\hfil
\subfloat[$N = 10$ antennas]{\includegraphics[width=2.5in]{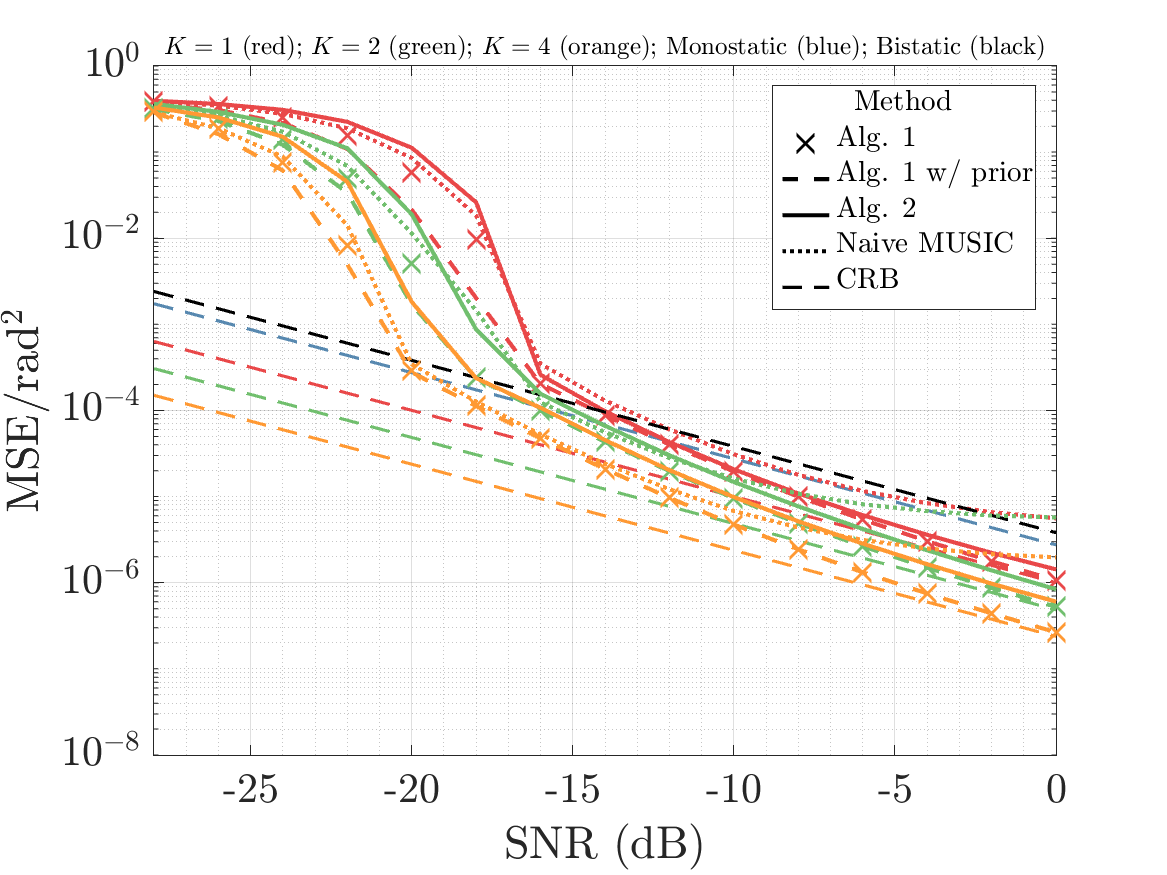}%
\label{fig:scene1_2_q3}}
\caption{Scenario 1: The \ac{MSE} versus \ac{SNR} of the proposed methods and benchmarks for a different number of users and $q=3$ targets located at $0^\circ,30^\circ,60^\circ$.}
\label{fig:scene1_q3}
\end{figure*}
\paragraph{Scenario $1$ with multiple targets}
In Fig. \ref{fig:scene1_q3}, we follow the same scenario as in Fig. \ref{fig:scene1_q1}, but this time tripling the number of targets, i.e. $q=3$. The targets are assumed to be located at $0^\circ,30^\circ,60^\circ$, respectively. This study will show the improvements of \ac{MSE} estimates with an increasing number of users, even with more targets. The number of subcarriers and \ac{OFDM} symbols for the $k^{th}$ user is also the same as that in Fig. \ref{fig:scene1_q1}. Note that the case of $K=0$ refers to the \textit{mono-static only} case, where only the echo received due to the \ac{DL} signal is used at the \ac{DFRC} \ac{BS}.
 In Fig. \ref{fig:scene1_1_q3}, we fix $N =5$ antennas. As was previously the case, we can observe that the benchmark of \textbf{Algorithm \ref{alg:alg1}} with \ac{AoA} prior achieves the best performance for a different number of users. The improvement of performance with increasing users is still preserved, as we see a factor of $3\dB$ gain in \ac{CRB}, when doubling $K$. Nevertheless, both versions of \textbf{Algorithm \ref{alg:alg1}} achieve the \ac{CRB} starting at $\SNR = 0 \dB$. Moreover, we see that \textbf{Algorithm \ref{alg:alg2}} generally outperforms the Naive \ac{MUSIC} benchmark starting from $\SNR =-11\dB$ for $K = 1$, $\SNR = -10\dB$ for $K = 8$, which can also be explained by the availability of more time resources, which becomes more notable for a higher number of users.
In Fig. \ref{fig:scene1_2_q3}, we fix $N = 10$ antennas, as well as $q=3$ targets. We can also say that the gap between both versions of \textbf{Algorithm \ref{alg:alg1}} shrinks with increasing $K$. For e.g., both versions coincide at an accuracy of $10^{-2}$ and for $K =16$, as compared to a $0.5\dB$ gap for $K = 1$. The Naive \ac{MUSIC} benchmark also approaches the \ac{CRB} for higher $N$, but saturates after $\SNR = 4\dB$ for $K = 1,2$. This loss of performance in Naive \ac{MUSIC} is due to the fact that it does not exploit the common manifold and, so, treats all \ac{AoAs}, whether targets or users, equally. We observe that for $N = 5$ antennas the \ac{CRB} for the \ac{SU} \ac{HRF} improves that of the mono-static by $5 \dB$, whereas there is a $10\dB$ gap between the mono-static and bi-static cases.
For $N = 10$, we see that the bi-static \ac{CRB} approaches that of the mono-static. However, the \ac{SU} \ac{HRF} improves by $4\dB$. 
\begin{figure*}[!t]
\centering
\subfloat[$N = 5$ antennas]{\includegraphics[width=2.5in]{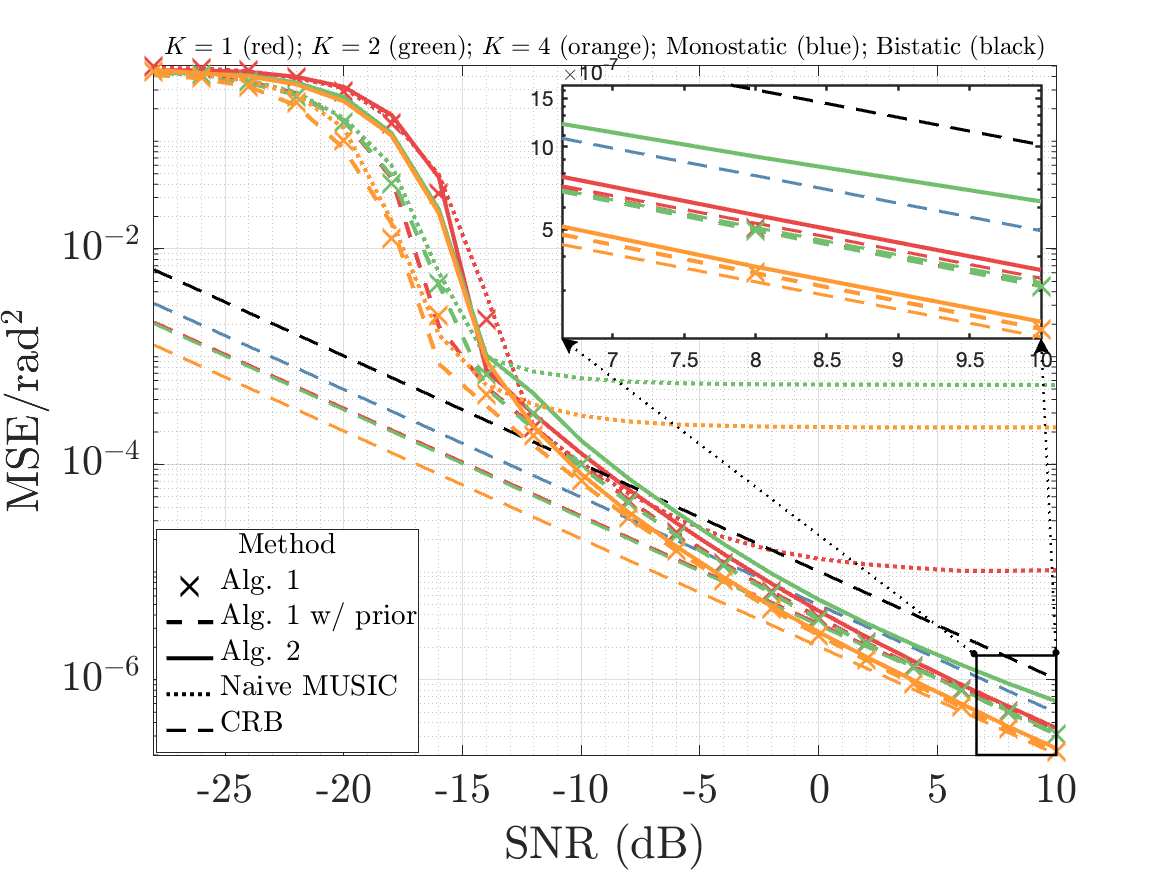}%
\label{fig:scene2_1_q1}}
\hfil
\subfloat[$N = 10$ antennas]{\includegraphics[width=2.5in]{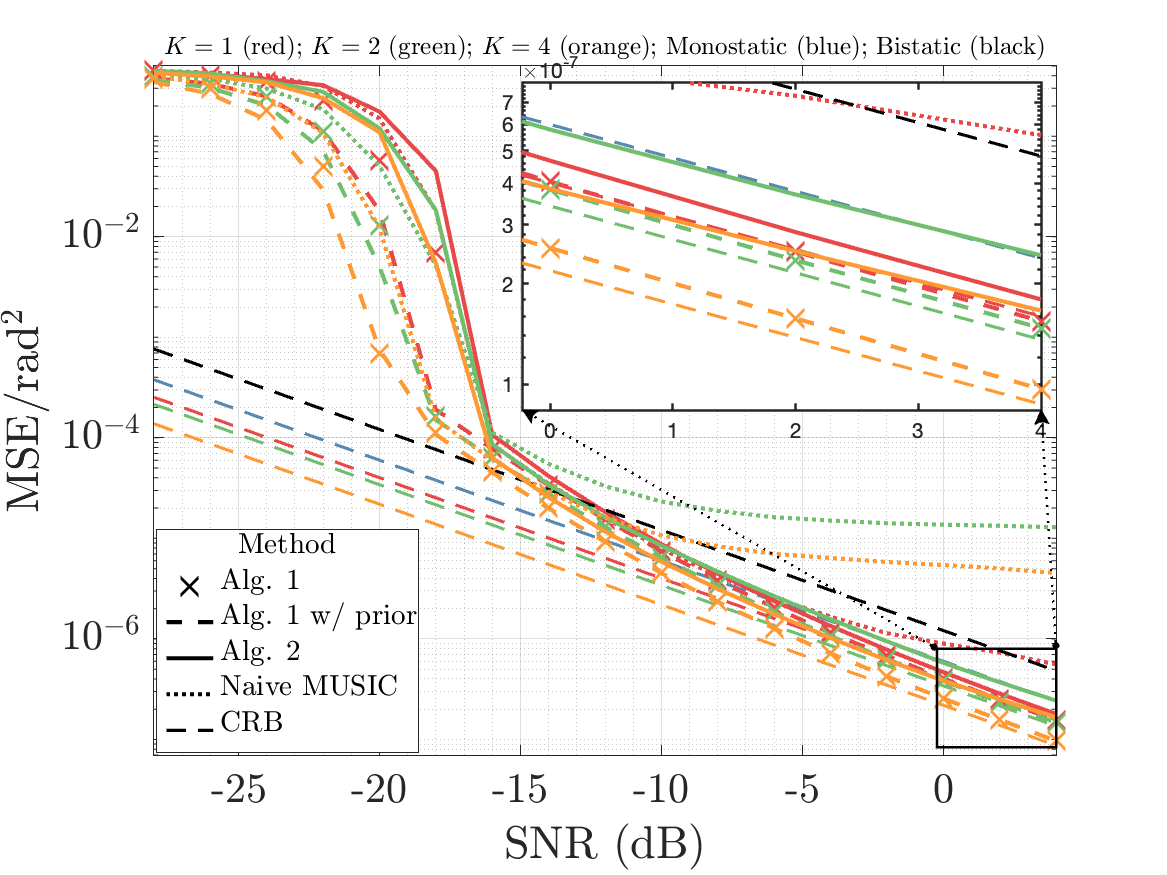}%
\label{fig:scene2_2_q1}}
\caption{Scenario 2: The \ac{MSE} versus \ac{SNR} of the proposed methods and benchmarks for different number of users and $q=1$ target located at $0^\circ$.}
\label{fig:scene2_q1}
\end{figure*}
\vspace{-0.1cm}
\paragraph{Scenario $2$ with a single target}
In Fig. \ref{fig:scene2_q1}, we switch to Scenario $2$, where we plot the \ac{MSE} of the estimated \ac{AoA}s as a function of varying \ac{SNR}, for different number of antennas. As done previously, we study the performance for $N=5$ and $N=10$ antennas.
In Fig. \ref{fig:scene2_1_q1}, we fix $N =5$ antennas for a single target located at $0^\circ$. As is the case in previous experiments, both versions of \textbf{Algorithm \ref{alg:alg1}} coincide and meet the \ac{CRB} at high \ac{SNR}. Also, \textbf{Algorithm \ref{alg:alg2}} decays with the \ac{CRB}, however, the \ac{MSE} of Naive \ac{MUSIC} saturates with increasing \ac{SNR}. For instance, focusing on $K=2$, we can see that at $\SNR = -14\dB$, the \ac{MSE} curve of Naive \ac{MUSIC} begins to saturate, whereas the \ac{MSE} of \textbf{Algorithm \ref{alg:alg2}} approaches the \ac{CRB} with an increase of \ac{SNR}. This is explained through \textbf{Algorithm \ref{alg:alg2}}'s effectiveness in segregating the common manifold. A similar observation is given for all other values of $K$. 
In Fig. \ref{fig:scene2_2_q1}, we fix $N = 10$ antennas with a single source as in Fig. \ref{fig:scene2_1_q1}. We can also report that both versions of \textbf{Algorithm \ref{alg:alg1}} converge towards the \ac{CRB} at high \ac{SNR}. Furthermore, for an \ac{MSE} level of $10^{-4} \rad^2$, there is a $0.4\dB$ gain when doubling the number of users from $K=1$ to $K=2$, followed by a gain of $0.6\dB$ between $K=2$ and $K=4$. This gain is explained via the availability of more bandwidth, where by increasing $K$, we also increase the bandwidth to accommodate users. Moreover, each user will provide a better spatial view of the target through its \ac{UL} contribution. We notice that the \ac{CRB} for the $K = 1$ case of \ac{HRF} improves the mono-static \ac{CRB} by $1.8\dB$, and $4.8\dB$ better than the bi-static configuration.  
\begin{figure*}[!t]
\centering
\subfloat[$N = 5$ antennas]{\includegraphics[width=2.5in]{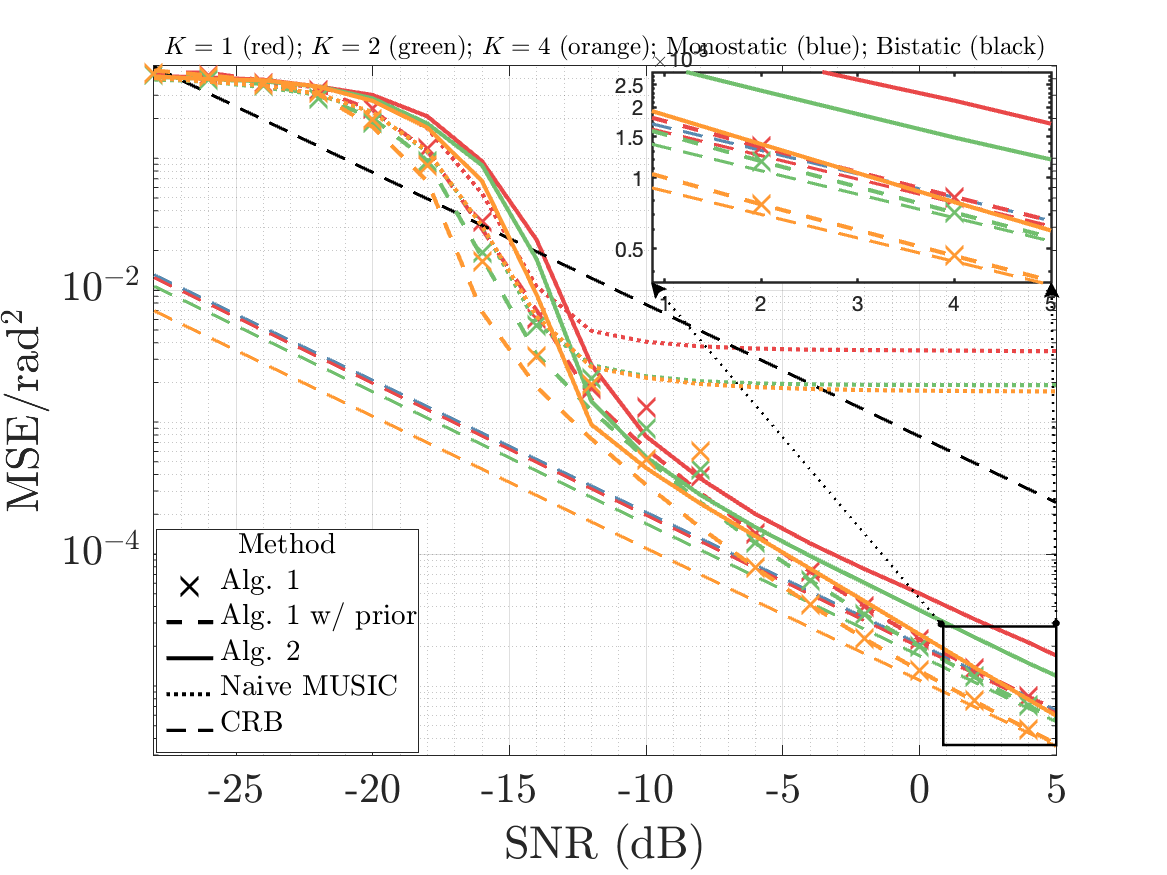}%
\label{fig:scene2_1_q3}}
\hfil
\subfloat[$N = 10$ antennas]{\includegraphics[width=2.5in]{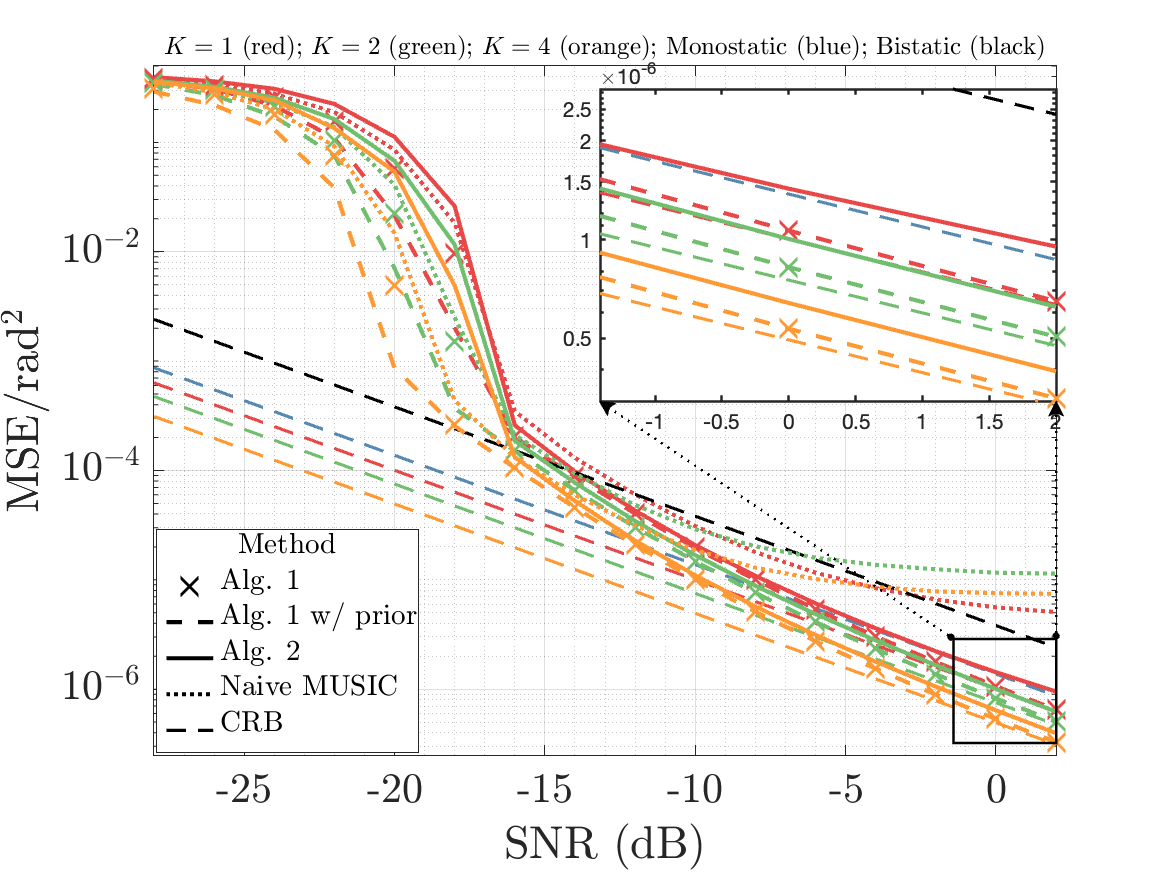}%
\label{fig:scene2_2_q3}}
\caption{Scenario 2: The \ac{MSE} versus \ac{SNR} of the proposed methods and benchmarks for a different number of users and $q=3$ targets located at $0^\circ,30^\circ,60^\circ$.}
\label{fig:scene2_q3}
\end{figure*}
\paragraph{Scenario $2$ with multiple targets}
In Fig. \ref{fig:scene2_q3}, we follow the same scenario as that in Fig. \ref{fig:scene2_q1}, but we place $q=3$ targets, instead of $q=1$. As is the case in Fig. \ref{fig:scene1_q3}, the targets are assumed to be located at $0^\circ,30^\circ,60^\circ$, respectively. We aim to study the performance with multiple targets.  
 In Fig. \ref{fig:scene2_1_q3}, we fix $N = 5$ and $q=3$. We observe that both versions of \textbf{Algorithm \ref{alg:alg1}} coincide with the \ac{CRB} as \ac{SNR} increases. Observe that, in general, Naive \ac{MUSIC} saturates earlier than the case of $q=1$, i.e. compared to Fig. \ref{fig:scene2_1_q1}. For e.g., at $K=1$, the Naive \ac{MUSIC} begins to saturate at $\SNR = -10\dB$ at an \ac{MSE} level of about $0.003\rad^2$, as compared to $10^{-5} \rad^2$ at $\SNR = 0 \dB$, in Fig. \ref{fig:scene2_1_q1}. This goes ahead to show that Naive \ac{MUSIC} is expected to do worse for more targets, due to its inability to distinguish the common target manifold. In contrast, \textbf{Algorithm \ref{alg:alg2}} approaches the \ac{CRB} for high \ac{SNR}.
In Fig. \ref{fig:scene2_2_q3}, we fix $N = 10$ and $q=3$. Again, we observe that both versions of \textbf{Algorithm \ref{alg:alg1}} converge to the \ac{CRB} at high \ac{SNR}, whereas the Naive \ac{MUSIC} saturates. \textbf{Algorithm \ref{alg:alg2}} decays with the \ac{CRB} for high \ac{SNR}. We also see gains for \textbf{Algorithm \ref{alg:alg2}}, when compared to Naive \ac{MUSIC}. For e.g., at an \ac{MSE} level of $10^{-5} \rad^2$ and at $K=1$, we see that \textbf{Algorithm \ref{alg:alg2}} gains about $2.5\dB$ compared to Naive \ac{MUSIC} and a $5\dB$ gain for $K=4$. Moreover, by referring to Fig. \ref{fig:scene2_2_q3}, we notice that the \ac{SU} \ac{HRF} \ac{CRB} is $1.3\dB$ better than the mono-static case, and also improves the bi-static case by $5\dB$. 
\begin{figure}[!t]
\centering
\includegraphics[width=2.5in]{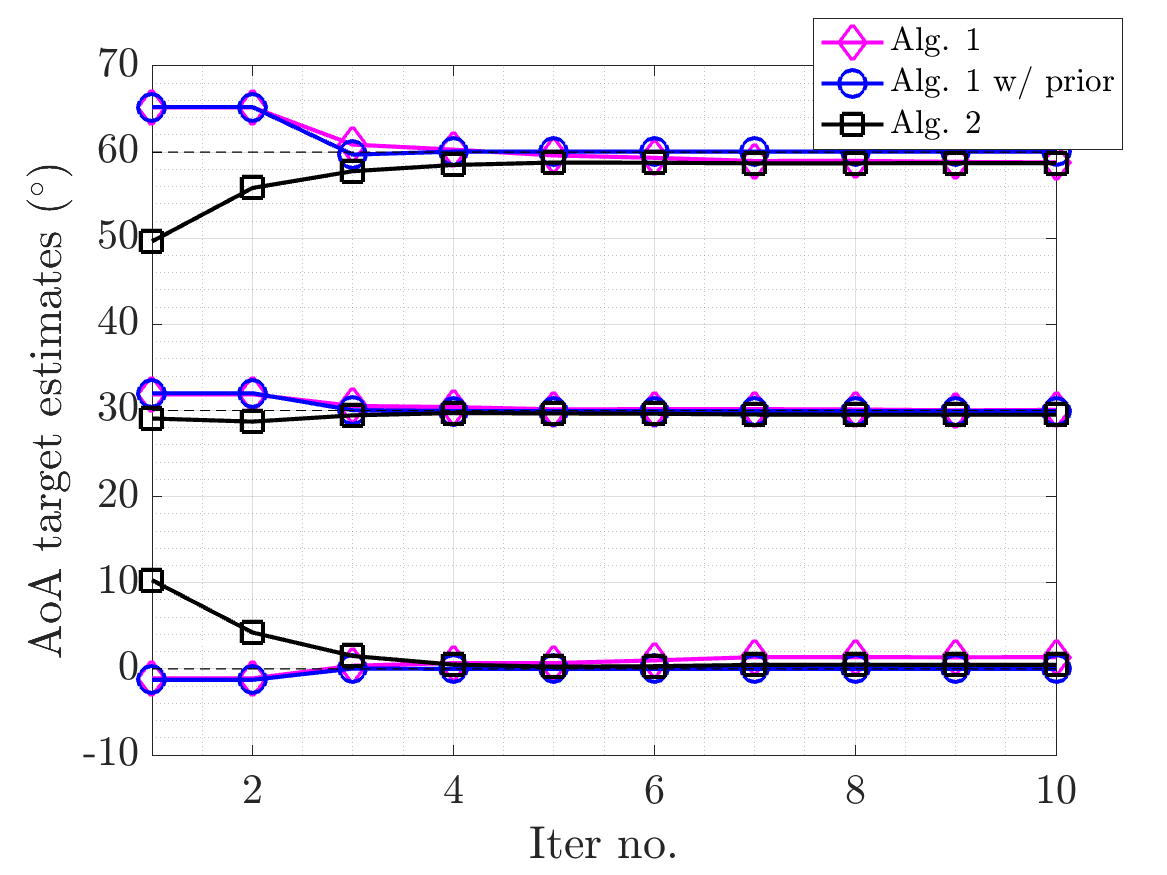}
\caption{Behavior of the different algorithms as a function of iteration number.}
\label{fig:iterations}
\end{figure}
\vspace{-0.3cm}
\paragraph{\ac{AoA} Behavior}
In Fig. \ref{fig:iterations}, we study the estimated \ac{AoA} behavior of the proposed methods as a function of iteration number. The estimate of each \ac{AoA} at any given iteration is averaged over Monte Carlo trials. The chosen $\SNR$ is $-12\dB$, $N=5$ and $K=16$. We observe that all methods converge, even at low $\SNR$. Even more, when the user \ac{AoAs} are fed as prior information into \textbf{Algorithm \ref{alg:alg1}}, we can see a slightly faster convergence than the case without prior information, most notably around the $3^{rd}$ iteration in an attempt to estimate $\theta_3^{\tt{tar}} = 60^{\circ}$. Also, all methods converge after $5$ iterations. This proves the effectiveness and stability of the proposed methods, from a convergence viewpoint, even at low $\SNR$. 
\begin{figure}[!t]
\centering
\includegraphics[width=2.5in]{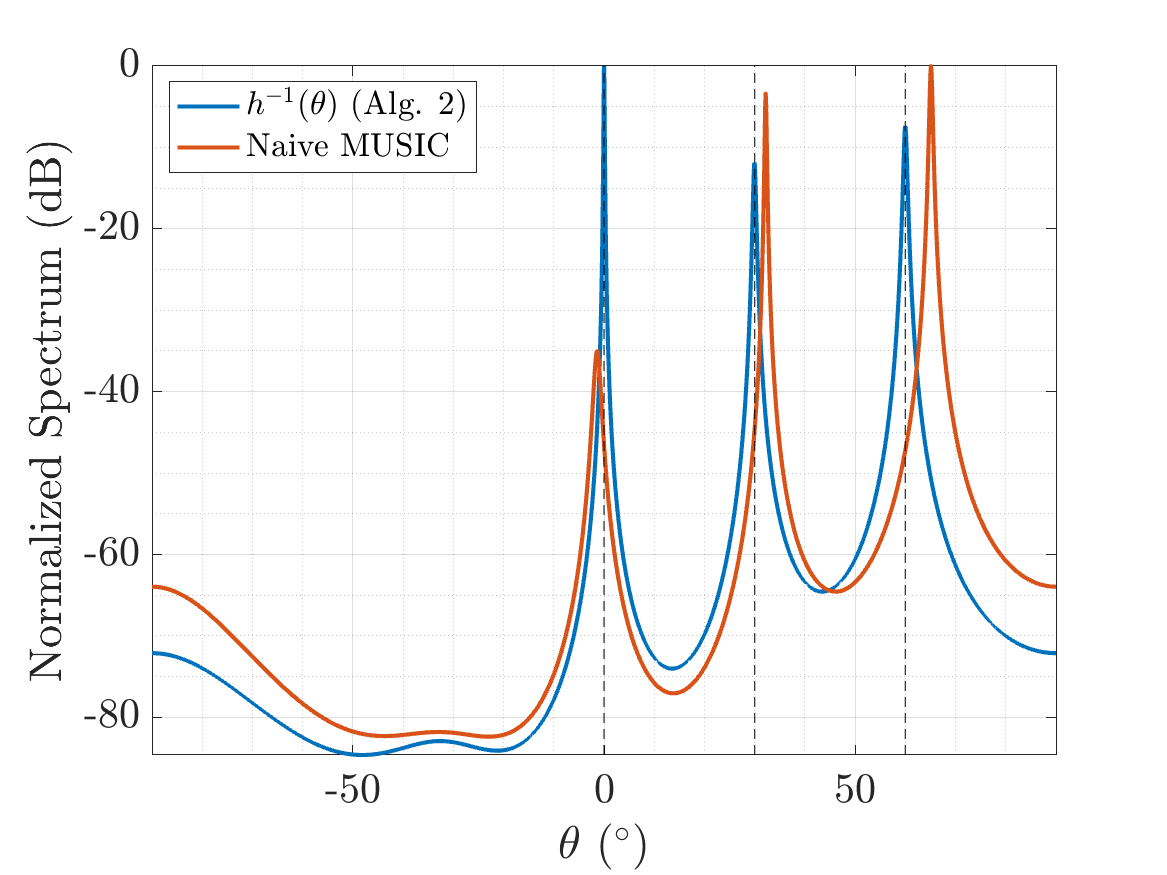}
\caption{A spectrum comparison of $h^{-1}(\theta)$ given in Algorithm \ref{alg:alg2} vs. the \ac{MUSIC} spectrum. Vertical dashed lines correspond to the true target \ac{AoAs}.}
\label{fig:spectrum}
\end{figure}
\paragraph{Spectra Comparison}
In Fig. \ref{fig:spectrum}, we reveal some internal dynamics of \textbf{Algorithm \ref{alg:alg2}}, namely the spectrum $h^{-1}(\theta)$, where its maximization is involved within the second phase of \textbf{Algorithm \ref{alg:alg2}}. So, a reasonable comparison of spectra would be $h^{-1}(\theta)$, compared to the classical \ac{MUSIC} spectrum. Besides, the spectra of each method have been averaged over multiple trials. Herein, we fix $\SNR = 10\dB$, $K = 16$ users, $q = 3$ located at $0^\circ,30^\circ,60^\circ$ and $N = 5$. Not only can we notice an accurate estimation of the true target \ac{AoAs}, but we can also report sharper peaks and an improved resolution, as compared to the Naive \ac{MUSIC} one, which further allows for better target \ac{AoA} estimation. Also, notice that the noise level of the spectra has been reduced by more than $6\dB$, especially around the region $[-70^\circ,-50^\circ]$. This again proves the superiority of \textbf{Algorithm \ref{alg:alg2}}, when compared to \ac{MUSIC}.
\begin{figure}[!t]
\centering
\includegraphics[width=3in]{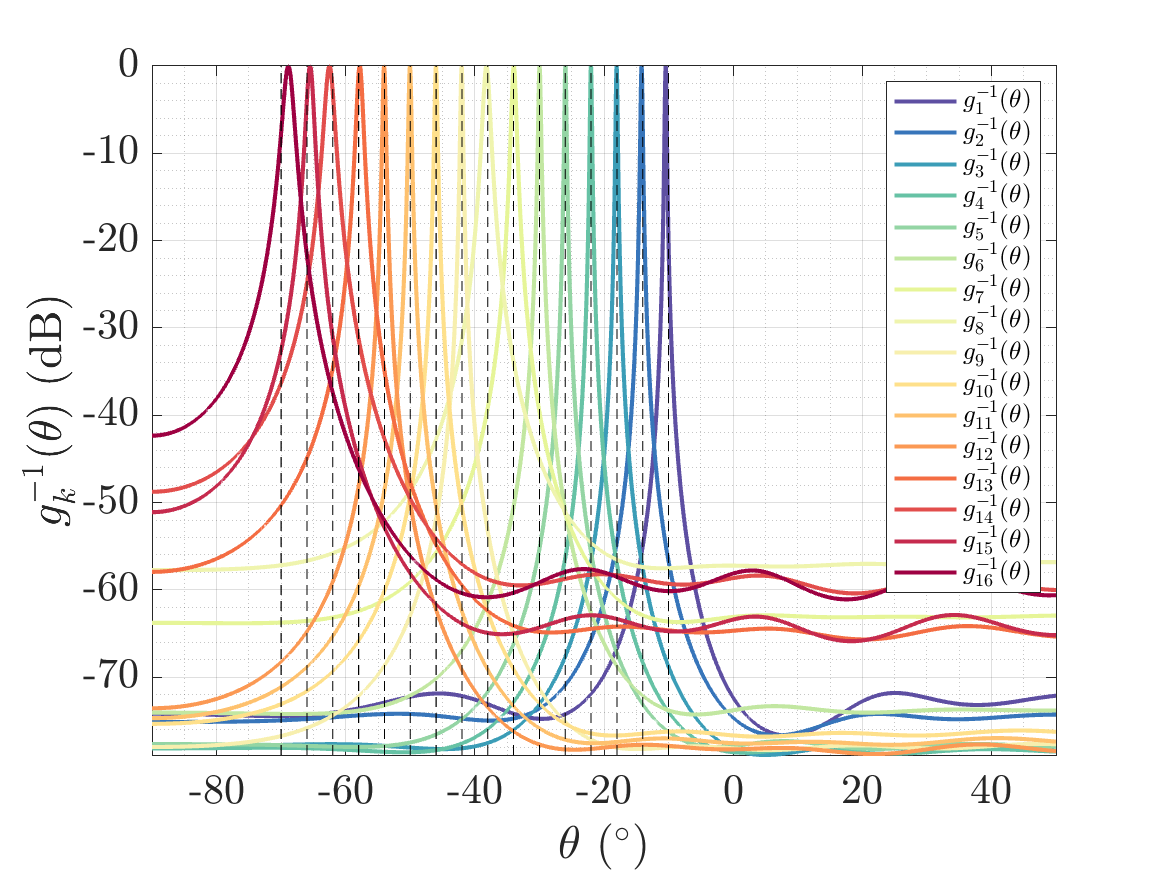}
\caption{The different spectra, i.e. $g_1^{-1}(\theta) \ldots g_K^{-1}(\theta)$, involved in the first phase of Algorithm \ref{alg:alg2}. Vertical dashed lines correspond to the true user \ac{AoAs}.}
\label{fig:spectrum_g}
\end{figure}
\paragraph{Behavior of $g(\theta)$}
In Fig. \ref{fig:spectrum_g}, we further highlight additional dynamics of \textbf{Algorithm \ref{alg:alg2}}, namely the spectrum $g_k^{-1}(\theta)$, found in the first phase of \textbf{Algorithm \ref{alg:alg2}}. As a reminder, the $k^{th}$ spectrum, $g_k^{-1}(\theta)$, aids in identifying the \ac{AoA} of the $k^{th}$ user. As is the case of Fig. \ref{fig:spectrum}, each spectrum is averaged over different Monte Carlo trials. The $\SNR$ is set to $10\dB$, the number of users is set to $K = 16$, and the number of antennas is set to $N =5$ antennas. First, we can see one and only one peak for any $g_k^{-1}(\theta)$. This is a good indicator, as $g_k^{-1}(\theta)$ should be highly selective of its associated user (i.e. the $k^{th}$ user) and also capable of simultaneously rejecting off all other users and targets, even with a low number of antennas i.e., $N=5$. With this feature in mind, we see that there are no spurious peaks arising in any of the spectra. Moreover, we can observe that the peaks sharpen, as the user approaches the broad-fire of the antenna array. Furthermore, we notice that the noise levels decrease as the user approaches the end-fire of the antenna array, which is a typical characteristic of \ac{ULA}.

\begin{figure}[!t]
\centering
\includegraphics[width=3in]{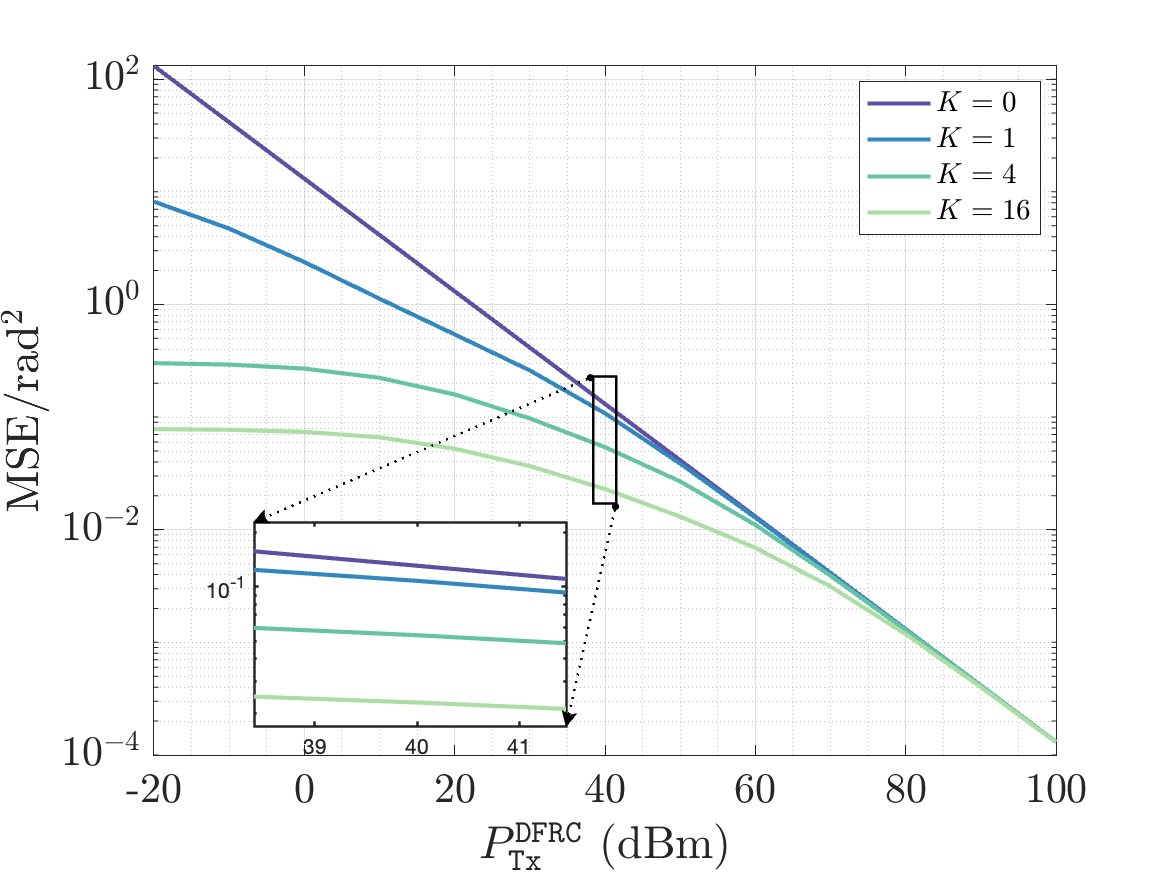}
\caption{The \ac{CRB} performance as a function of transmit \ac{DFRC} \ac{BS} power for a typical \ac{mmWave} communication system.}
\label{fig:CRB_vs_PTxDFRC}
\end{figure}

\begin{figure}[!t]
\centering
\includegraphics[width=3in]{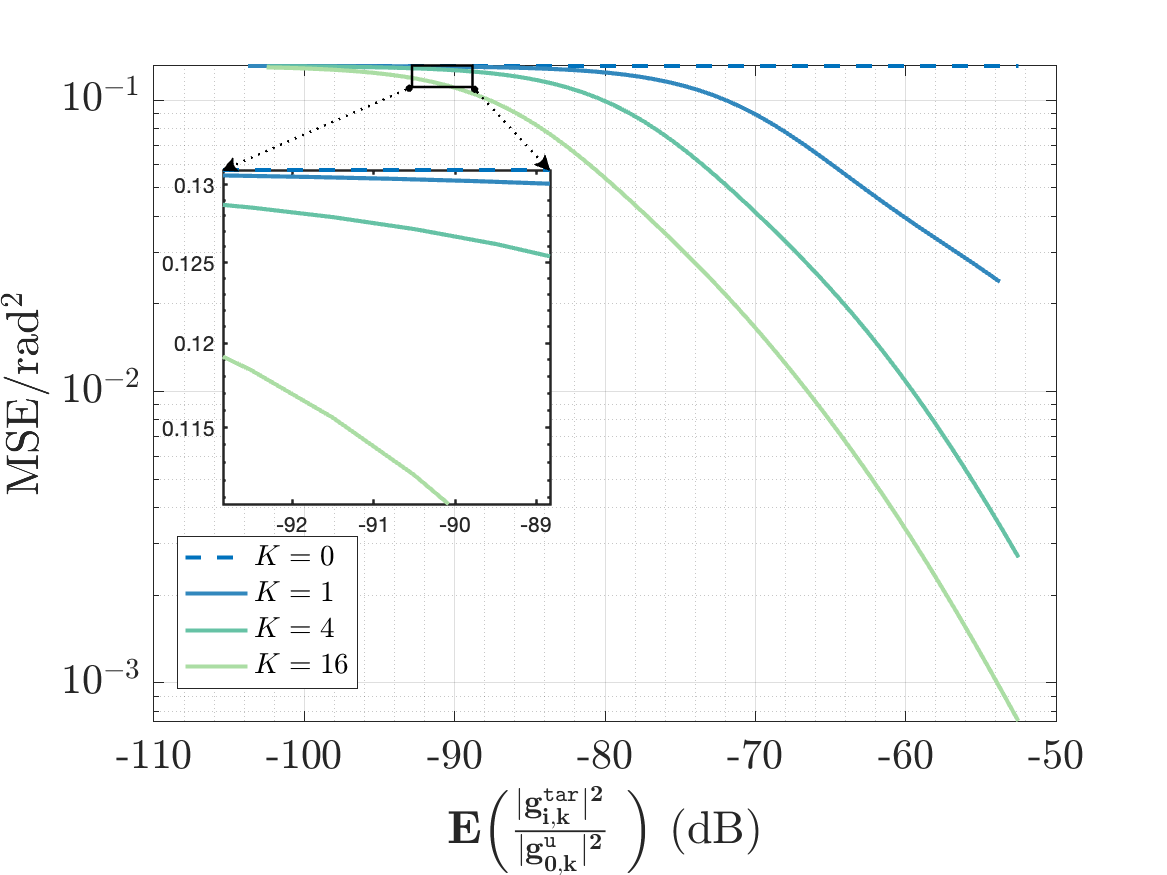}
\caption{The \ac{CRB} performance as a function of mean reflection-to-\ac{LoS} power for a typical \ac{mmWave} communication system at $P_{\tt{Tx}}^{\tt{DFRC}} = 40\dBm$.}
\label{fig:CRB_vs_PLRatio}
\end{figure}
\vspace{-0.05cm}

\paragraph{\ac{HRF} Necessity}
\label{paragraph:hrf-necessity}
In practical scenarios, the communication user is characterized by a notably lower transmission power and deploys a comparatively smaller antenna array in contrast to the infrastructure of the \ac{DFRC} \ac{BS} \cite{8207426}.
The purpose of the simulation depicted in Fig. \ref{fig:CRB_vs_PTxDFRC} is to examine the conditions where the utilization of the \ac{HRF} becomes imperative for enhancing sensing performance across a range of diverse communication users.
For that, we have utilized the same simulation parameters as those for a typical \ac{mmWave} communication system \cite{8207426} (c.f. Table XIV), and we have set the total power transmitted by the \ac{DFRC} \ac{BS} as $P_{\tt{Tx}}^{\tt{DFRC}}$ to study the attainable \ac{CRB} for various number of \ac{UL} users.
We notice that at $P_{\tt{Tx}}^{\tt{DFRC}} = 40\dBm$, which is the transmit power of a \ac{BS} in typical \ac{mmWave} systems \cite{8207426}, the \ac{HRF} continues to bestow tangible sensing advantages with increasing number of users. 
For example, when $K = 0$ the \ac{CRB} is $0.13 \rad^2$ and with the advent of $K = 1$, the \ac{CRB} experiences a reduction to $0.1\rad^2$. Upon quadrupling the user count, a proportionate reduction factor of $2$ emerges, resulting in \ac{CRB} values of $0.05 \rad^2$ for $K = 4$ and $0.025\rad^2$ for $K = 16$.
Another noteworthy aspect to consider is the threshold transmit power level at which the \ac{HRF} may no longer be useful.
At $52\dBm$, it becomes evident that the performance of a single user ($K = 1$) in the \ac{HRF} scheme aligns closely with the performance of the \ac{DFRC} \ac{BS} operating in isolation.
Similarly, when the transmit power exceeds $66\dBm$, we observe that the performance of the \ac{HRF} scheme with $K = 4$ coincides with that of the \ac{DFRC} \ac{BS} only-case.
Moreover, upon reaching a transmit power of $78\dBm$, the convergence persists, with $K = 16$ users in the \ac{HRF} scheme closely mirroring the performance of the \ac{DFRC} \ac{BS} in standalone operation.
The insight derived from these observations suggests that, in scenarios where transmit power surpasses the specified thresholds, the deployment of the \ac{HRF} scheme may not yield substantial additional sensing gains beyond the capabilities of the \ac{DFRC} operating independently.
In Fig. \ref{fig:CRB_vs_PLRatio}, we aim to study the \ac{CRB} evolution as a function of the ratio of the reflected power to the direct \ac{LoS} transmission power. 
In this simulation, we fix $P_{\tt{Tx}}^{\tt{DFRC}} = 40\dBm$ as specified in \cite{8207426}.
In the general case, our simulations reveal that the advantages of \ac{HRF} remain significant, particularly when reflection-to-\ac{LoS} power ratio is reasonable, and for increasing $K$.
When the ratio between reflection-to-\ac{LoS} power ratio falls below the threshold of $-90\dB$, the performance for $K = 1$ users aligns with that of the standalone \ac{DFRC} \ac{BS}. 
Similarly, as the reflection-to-\ac{LoS} power ratio decreases to values below $-96\dB$ and $-102\dB$, the performance for $K = 4$ and $K = 16$ users respectively coincides with that of the \ac{DFRC} \ac{BS} operating on its own.
The observations suggest that as the reflection-to-\ac{LoS} ratio decreases below certain thresholds, the performance of multiple users contributing to \ac{HRF} approaches the performance level exhibited by the standalone \ac{DFRC} \ac{BS}.
In addition, these thresholds decrease with increasing number of users participating for \ac{HRF} operations. 
This insight underscores the significance of the reflection-to-LoS ratio in determining the system's capability to support multiple users and indicates a critical threshold where the system's performance is notably affected by the ratio of reflected to direct signals.
In general, we see that quadrupling the number of users requires $6\dB$ to $7\dB$ less reflection power to achieve the same \ac{CRB} performance.

\vspace{-0.35cm}
\section{Conclusions and Outlook}
\label{sec:conclusions}
In this paper, a new hybrid radar model is proposed, where the \ac{DFRC} \ac{BS} acts as a monostatic radar in the \ac{DL} and users act as distributed bistatic radars in the \ac{UL}, following an \ac{FDD} protocol. From the \ac{DFRC} perspective, we derive the \ac{ML} criterion of the \ac{AoA} estimation problem, and assemble its connections with the classical \ac{AoA} estimation problem. We take a step further in proposing two efficient estimators. The first one is built on the direct optimization of the \ac{FML} criterion by alternating between user and target terms in the \ac{ML} cost function. Furthermore, the second method diligently forms an iterative procedure via the formulation of suitable optimization methods that exploit the common manifold structure appearing over different communication bands. Our simulation findings prove the effectiveness of the proposed hybrid radar model, along with methods involved in resolving and estimating the locations. In short, the proposed communication-centric hybrid radar model has proven its potential and superiority in improving target location estimates in different settings. 

Research directions include extensions toward wide-band, which further lead to frequency-dependent array response vectors. Also, due to full duplex operation, we plan to propose estimators, and compensators, for the self-interference leakage phenomenon caused by the transmitting onto the receipting unit of the \ac{DFRC} \ac{BS}. Another direction is the multi-parameter estimation case, where the \ac{ToA} and Doppler parameters are to be jointly estimated with \ac{AoAs}. Moreover, from an estimation perspective, we also allude to the need for more sophisticated algorithms that trade complexity for performance.


\bibliographystyle{IEEEtran}
\bibliography{refs}

\vfill

\end{document}